\documentclass[bibyear]{aa}
\usepackage{graphicx}
\usepackage{txfonts}
\usepackage{hyperref}
\usepackage{mathabx}
\usepackage{natbib}
\usepackage{tabularx}
\usepackage{multirow}
\usepackage{multicol, blindtext}
\usepackage{color}
\usepackage{amsmath}
\usepackage{xcolor}
\usepackage[toc,title,page]{appendix} 
\usepackage{geometry}
\usepackage{arydshln}
\usepackage{amssymb}
\usepackage{ulem}
\DeclareMathOperator{\sign}{sign}

\newcommand{\Teffs}{{T_{\mathrm{eff,\star}}}}
\newcommand{\BJDTDB}{\mathrm{BJD}_\mathrm{TDB}}
\newcommand{\BJDUTC}{\mathrm{BJD}_\mathrm{UTC}}
\newcommand{\Porb}{P_{\mathrm{orb}}}
\newcommand{\Prots}{P_\mathrm{rot,\star}}
\newcommand{\Protp}{P_\mathrm{rot,p}}
\newcommand{\rp}{r_\mathrm{p}}
\newcommand{\mpl}{m_\mathrm{p}}
\newcommand{\Rs}{R_\mathrm{\star}}
\newcommand{\Ms}{M_\mathrm{\star}}
\newcommand{\esin}{\sqrt{e}\cdot \sin \omega_0}
\newcommand{\ecos}{\sqrt{e}\cdot \cos \omega_0}
\newcommand{\Loves}{k_\mathrm{{2,\star}}}
\newcommand{\Lovep}{k_\mathrm{{2,p}}}

\begin{document}



   \title{Evidence of apsidal motion and a possible co-moving companion star detected in the WASP-19 system}

   \author{L. M. Bernab\`o \inst{\ref{inst:PF@DLR}}$^{\href{https://orcid.org/0000-0002-8035-1032}{\includegraphics[scale=0.5]{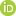}}}$
          \and Sz. Csizmadia \inst{\ref{inst:PF@DLR},\ref{inst:ELKH}}$^{\href{https://orcid.org/0000-0001-6803-9698}{\includegraphics[scale=0.5]{Images/orcid.jpg}}}$
          \and A.~M.~S.~Smith\inst{\ref{inst:PF@DLR}}$^{\href{https://orcid.org/0000-0002-2386-4341}{\includegraphics[scale=0.5]{Images/orcid.jpg}}}$
          \and H. Rauer\inst{\ref{inst:PF@DLR},\ref{inst:ZAA},\ref{inst:FU}}$^{\href{https://orcid.org/0000-0002-6510-1828}{\includegraphics[scale=0.5]{Images/orcid.jpg}}}$
          \and A. Hatzes\inst{\ref{inst:TLS}}$^{\href{https://orcid.org/0000-0002-3404-8358}{\includegraphics[scale=0.5]{Images/orcid.jpg}}}$
          \and M. Esposito\inst{\ref{inst:TLS}}$^{\href{https://orcid.org/0000-0002-6893-4534}{\includegraphics[scale=0.5]{Images/orcid.jpg}}}$
          \and D. Gandolfi\inst{\ref{inst:Torino}}$^{\href{https://orcid.org/0000-0001-8627-9628}{\includegraphics[scale=0.5]{Images/orcid.jpg}}}$
          \and J. Cabrera\inst{\ref{inst:PF@DLR}}$^{\href{https://orcid.org/0000-0001-6653-5487}{\includegraphics[scale=0.5]{Images/orcid.jpg}}}$ 
          }

   \institute{
   \label{inst:PF@DLR}            Institute of Planetary Research, German Aerospace Center (DLR), Rutherfordstrasse 2, 12489 Berlin
   \and \label{inst:ELKH} ELKH-SZTE Stellar Astrophysics Research Group, H-6500 Baja, Szegedi \'ut Kt. 766, Hungary
   \and \label{inst:ZAA}          Zentrum f{\"u}r Astronomie und Astrophysik, Technische Universit{\"a}t Berlin, Hardenbergstr. 36, D-10623 Berlin, Germany
   \and \label{inst:FU}           Institut f{\"u}r Geologische Wissenschaften, Freie Universit{\"a}t Berlin, 12249 Berlin, Germany
   \and \label{inst:TLS}          Th{\"u}ringer Landessternwarte Tautenburg, Sternwarte 5, 07778 Tautenburg, Germany
   \and \label{inst:Torino}       Dipartimento di Fisica, Universit{\`a} degli Studi di Torino, Via Pietro Giuria, 1, 10125 Torino, Italy\\
   \email{lia.bernabo@dlr.de}\\
   }

 \date{Submitted 9 May 2023 ; accepted 8 January 2024}

 \abstract
   {Love numbers measure the reaction of a celestial body to perturbing forces, such as the centrifugal force caused by rotation, or tidal forces resulting from the interaction with a companion body. These parameters are related to the interior density profile. The non-point mass nature of the host star and a planet orbiting around each other contributes to the periastron precession. The rate of this precession is characterized mainly by the second-order Love number, which offers an opportunity to determine its value. When it is known, the planetary interior structure can be studied with one additional constraint beyond the mass, radius, and orbital parameters.}
   {We aim to re-determine the orbital period, eccentricity, and argument of the periastron for WASP-19Ab, along with a study of its periastron precession rate. We calculated the planetary Love number from the observed periastron precession rate, based on the assumption of the stellar Love number from stellar evolutionary models.}
   {We collected all available radial velocity (RV) data, along with the transit and occultation times from the previous investigations of the system. We supplemented the data set with 19 new RV data points of the host star WASP-19A obtained by HARPS. Here, we summarize the technique for modeling the RV observations and the photometric transit timing variations (TTVs) to determine the rate of periastron precession in this system for the first time.}
   {We excluded the presence of a second possible planet up to a period of $\sim$4200 d and with a radial velocity amplitude bigger than $\simeq$1 m s$^{-1}$. We show that a constant period is not able to reproduce the observed radial velocities. We also investigated and excluded the possibility of tidal decay and long-term acceleration in the system. However, the inclusion of a small periastron precession term did indeed improve the quality of the fit. We measured the periastron precession rate to be 233 $^{+25}_{-35}$ $^{\prime\prime} d^{-1}$. By assuming synchronous rotation for the planet, it indicates a $k_2$ Love number of 0.20 $^{+0.02}_{-0.03}$ for WASP-19Ab.
   }
   {The derived $\Lovep$ value of the planet has the same order of magnitude as the estimated fluid Love number of other Jupiter-sized exoplanets (WASP-18Ab, WASP-103b, and WASP-121b). A low value of $\Lovep$ indicates a higher concentration of mass toward the planetary nucleus. 
   }

   \keywords{}

  \titlerunning{Periastron precession of WASP-19Ab}
  \authorrunning{Bernabò et~al.}
   \maketitle
\section{Introduction}
The understanding of planetary interiors constitutes one of the four major challenges of contemporary exoplanetary science \citep{Schneider2018}. It is crucial for assessing their potential habitability, formation, and evolution. Interior processes play an essential role in creating and maintaining the physical conditions that are required to support life (see, e.g.,~\citealt{VanHoolst2019}). 
They are useful for determining which planetary formation process is most plausible; namely, whether it is core accretion (\citealt{Pollack1996} and references therein) or gravitational instability (\citealt{Boss1997} and references therein). According to the former, the accretion of planetesimals would result in the formation of a core comprised of heavy elements. 
On the other hand, planets forming through disc instability can rapidly accrete gas.
These two scenarios lead to different core sizes. The size of the core is related to the second order fluid $\Lovep$ Love number (\citealt{Love1911}), as explained in \citet{becker13}. Knowledge of the interior of a giant planet can help us distinguish between
these two scenarios, in addition to its Love number, as we describe in this work. 
\\
\indent Love numbers can also be used to break the known mass-radius-composition degeneracy of exoplanets \citep{Baumeister2020}. When an exoplanet is characterizable both by transit and radial velocity techniques, we can derive its radius and mass -- and, thus, the planetary mean density. However, mass, radius and mean density are degenerate when it comes to determining the interior structure of bodies, such as the radial density, temperature, pressure, and composition profiles. We can find multiple solutions for the planetary interior producing the same planetary total mass and radius (see, e.g., \citealt{valencia07,wagner11,damasso18}). This kind of degeneracy can be decreased based on the strong assumption that the planet and its host star share the same metallicity \citep{dorn15}.\\
\indent For Solar System planets, the degeneracy between mean density and interior structure can be reduced by in situ measurements of the gravitational field, as well as by seismic measurements, such as InSight (\citealt{Banerdt2020}), and by observations of moon motions. However, for exoplanets, gravitational moments cannot be measured. Hence, Love numbers were proposed as further observables in exoplanetary interior studies (see, e.g., \citealt{Batygin2009,RagozzineWolf2009} and references therein), as has already been applied to eclipsing binary stars systems (see~\citealt{Russell1928} and references therein). 
These authors measured the deformations and mass re-distribution inside the planet due to the tidal interaction with the host star. As shown by~\cite{Baumeister2020}, thanks to the knowledge of the second-order fluid Love number of the planet, we can better infer the distribution of possible thickness of each interior layer. 

\indent In this work, we employ a technique used by~\cite{Csizmadia2019}, which we further refined to determine $\Lovep$ from periastron precession. This precession is derived from radial velocity (RV) measurements and transit and occultation mid-times. We apply it to the system WASP-19 based on an approach that requires the calculation of the secular evolution of the orbital elements, which is related to the mass distribution of the star and the planet through their second-degree Love numbers.\\


\noindent This paper is organized as follows. Section~\ref{sec:orbital_elements} and Appendix~\ref{app:apsidal_motion} present the theory of apsidal motion. Section~\ref{sec:system} describes the system WASP-19 and provides evidence of a possible companion star in the WASP-19 system. In Sect.~\ref{sec:data}, we describe archival transit and occultation timing observations as well as the archival radial velocity investigations of the system which are supplemented by our recent RV study of the system. Section~\ref{sec:data_analysis} contains our data analysis along with the search for a second planetary companion. Our conclusions are given in Sect.~\ref{sec:discussion}.

\section{Apsidal motion}
\label{sec:orbital_elements}

Apsidal motion, namely, the secular perturbation in the argument of periastron, can be detected either in RV or in TTV datasets (or their combination) to constrain the internal structure of planets. 
This type of analysis on RVs was introduced by \citet{Kopal1959, Kopal1978}. Here, we make use of such equations, however, we have rewritten them in a way that allows them to be directly applied to the observed data sets. We describe their derivations more in detail in Appendix~\ref{app:apsidal_motion}. \\
\indent The total rate of apsidal advance is the result of the three contributions:
\begin{equation}
    \dot{\omega} = \frac{d\omega}{dt} = \frac{d\omega_\mathrm{GR}}{dt} + \frac{d\omega_\mathrm{tidal}}{dt} + \frac{d\omega_\mathrm{rot}}{dt} .
    \label{eq:wdot_tot}
\end{equation}
\indent The general relativistic term $d\omega_\mathrm{GR}/dt$ (\citealt{Einstein1915}) is:
\begin{equation}
    \frac{d\omega_\mathrm{GR}}{dt} = \frac{6 \pi G M_{\star}}{a c^2 (1-e^2)}\frac{1}{P_\mathrm{a}},
\end{equation}
where {\it G} is the gravitational constant, $\Ms$ the mass of the star, {\it a} and {\it e} the semi-major axis and eccentricity of the orbit, {\it c} the speed of light in vacuum, and $P_\mathrm{a}$ the anomalistic period. \\
\indent The secular perturbations of the tidal and rotational components $d\omega_\mathrm{tidal}/dt$ and $d\omega_\mathrm{rot}/dt$ are derived in Appendix~\ref{app:apsidal_motion} and can be expressed more explicitly if approximated to the second order, as in~\cite{Csizmadia2019}:
\begin{equation}
\begin{split}
    \frac{d\omega_\mathrm{tidal}}{dt} \simeq \frac{15}{{2}} \frac{n}{(1-e^2)^5} \left[ \frac{\mpl}{\Ms} \Loves \left(\frac{\Rs}{a}\right)^5 + \right. \\ + \left. \frac{\Ms}{\mpl} \Lovep \left(\frac{R_\mathrm{p}}{a}\right)^5  \right] \left( 1 + \frac{3}{2}e^2+ \frac{1}{8} e^4 \right),
    \label{eq:omega_dot_tidal}
\end{split}
\end{equation}
and
\begin{align}
\begin{split}
    \frac{d\omega_\mathrm{rot}}{dt} \simeq 
    \frac{1}{{2}}
    \frac{n}{(1-e^2)^2} \left[ \left( \frac{P_\mathrm{orb}}{P_\mathrm{rot,\star}} \right)^2 \Loves \left( \frac{R_\mathrm{\star}}{a}\right)^5 \left( 1 + \frac{\mpl}{\Ms} \right)+ \right. \\  \hspace{1.1cm} + \left. \left(\frac{\Porb}{P_\mathrm{rot,p}} \right)^2 \Lovep \left( \frac{R_\mathrm{p}}{a} \right)^5 \left(1+  \frac{\Ms}{\mpl} \right) \right].
    \label{eq:omega_dot_rot}
    \end{split}
\end{align}
\indent In these equations, $n$ is the mean motion, $\mpl$ and $R_\mathrm{p}$ the mass and radius of the planet, $\Rs$ the radius of the star, and $\Protp$ and $\Prots$ are the rotational periods of the two components. Also, $\Lovep$ and $\Loves$ are their second-order fluid Love number corresponding to the double the apsidal motion constant (\citep{Csizmadia2019}):
\begin{equation}
    k_\mathrm{2,i} = 2 k_\mathrm{2,aps,i},
\end{equation}
both for the star ($i=\star$) and the planet ($i=p$). These two equations were obtained by~\cite{Sterne1939}. Just as in Eq.~(\ref{eq:domega/dt}), it is assumed that the rotational axis of both components are perpendicular to the orbital plane.\\
\indent Regardless of whether the stellar and planetary rotation is direct or retrograde, the corresponding terms in Eq.~(\ref{eq:omega_dot_rot}) are positive, therefore they cause the apsidal line to advance. \\

\noindent Assuming the values of the orbital parameters of WASP-19Ab from literature and the Love number of the star from~\cite{Claret2004}, we calculated the amplitude of the general relativity term and the stellar tidal and rotational contributions, as reported in Table~\ref{tab:wdot_contributions}. 
Moreover, a first estimate for the planet can be calculated assuming synchronous rotation for the planet and $\Lovep$ between the values 0.01-1.5 (corresponding to the extreme cases of a mass point and a homogeneous body, respectively). 
The total contribution is between about 14.71 - 1689 $^{\prime\prime}$d$^{-1} \simeq$ 0.0041 - 0.479 $^{\circ}$ d$^{-1}$, respectively when assuming $\Lovep$=0.01 and $\Lovep$=1.5. The apsidal motion period $U$=2$\pi / \dot{\omega}$ is therefore between $\simeq$ 2.10 yrs and $\simeq$ 241.25 yrs. \\ 
\begin{table}[]
    \renewcommand{\arraystretch}{1.4}
    \centering
    \begin{tabular}{c | c c}
     & $\Lovep = 0.01$ & $\Lovep = 1.5$ \\ 
    \hline\hline 
    $\dot{\omega}_{GR}$ & 2.85 $^{\prime\prime}$ d$^{-1}$ & 2.85 $^{\prime\prime}$ d$^{-1}$ \\
    $\dot{\omega}_{tid, \star}$ & 0.47 $^{\prime\prime}$ d$^{-1}$ & 0.47  $^{\prime\prime}$ d$^{-1}$ \\
    $\dot{\omega}_{rot, \star}$ & 0.15 $^{\prime\prime}$ d$^{-1}$ & 0.15 $^{\prime\prime}$ d$^{-1}$\\
    $\dot{\omega}_{tid, p}$ & 10.53 $^{\prime\prime}$ d$^{-1}$ & 1580.16 $^{\prime\prime}$ d$^{-1}$ \\
    $\dot{\omega}_{rot, p}$ & 0.70 $^{\prime\prime}$ d$^{-1}$ & 105.46 $^{\prime\prime}$ d$^{-1}$ \\
    \hline
    Total $\dot{\omega}$ & 14.70 $^{\prime\prime}$ d$^{-1}$ & 1689.10 $^{\prime\prime}$ d$^{-1}$\\
    Period & 241.25 yrs & 2.10 yrs \\
    \end{tabular}
    \caption{Expected gravitational, rotational and tidal contribution to the periastron precession in the extreme Love number cases (mass-point and homogeneous planet) and the corresponding apsidal motion period, assuming synchronous rotation for the planet.}
    \label{tab:wdot_contributions}
\end{table}

The equations in Appendix~\ref{app:apsidal_motion} show that the angle of periastron $\omega$ oscillates over one orbital period but also shows a gradual long-term change that can be retrieved in the RV observations. Due to the finite length of the typical datasets of observations (12 years in the case of WASP-19Ab), in RV datasets, the change in the angle of periastron $\omega$ can be retrieved in the radial velocity of the star (see, e.g.,~\citealt{Jackson2008} for more details on the timescales). In long baselines of observations, all orbital elements must be considered along with periastron precession. As described in~\cite{Kopal1978}, they include the change in semi-major axis (tidal decay), eccentricity (circularization) and time of periastron passage. The semi-major axis and its eccentricity would diminish in value and the time of periastron passage is shifted in time. \\


\section{WASP-19Ab}
\label{sec:system}

\subsection{The system}
\label{subsec:systembackground}
The planet WASP-19Ab, hosted by the bright, active G-dwarf star ($V$=12.3, \citealt{Vizier2012} and $\Teffs\simeq$5568 K,~\citealt{Torres2012}), was discovered through the transit method by the Wide Angle Search for Planets (WASP)-South observatory (\citealt{Hebb2010}). It is a hot Jupiter with mass $m_\mathrm{p}$ = 1.139 $M_\mathrm{\Jupiter}$, inflated radius $r_\mathrm{p}$ = 1.410 $R_{\Jupiter}$ (\citealt{Mancini2013}) and orbital period $\Porb \simeq$ 0.79 d.\\ 
\indent So far, the system is among the best studied from the point of view of orbital and planetary parameters \citep[e.g.,][]{Lendl2013, Mancini2013, Sedaghati2015, Sedaghati2017} and atmosphere characterization (very pronounced water absorption,~\citealt{Huitson2013, Sedaghati2017}, and titanium oxide detected in the transmission spectrum,~\citealt{Sedaghati2017}). The proximity of WASP-19Ab to its host star results in significant tidal interaction that may govern the evolution of the planetary orbit, which makes the system a good candidate for our study.\\
\indent Based on transit observations, certain studies have suggested the possibility of variations in the mid-transit time.~\cite{Mancini2013} suggested a non-linear ephemeris. However,~\cite{Petrucci2020} discarded orbital decay with 74 complete transit light curves spanning over ten years of observations. Moreover, as described more in detail in Sect.~\ref{subsec:long_term},~\cite{CortesZuleta2020} used transit timing variations (TTVs) to infer an upper mass limit for an additional candidate planetary companion in the system.


\subsection{Companion star: WASP-19B}
\label{subsec:WASP-19B}

We searched the catalogue of the third {\it Gaia} data release (DR3) \citep{GAIA,GAIADR3} for possible co-moving companions to WASP-19A. We searched all objects within $4^\prime$ of the target, which corresponds to a physical separation of around 0.3~pc at the distance of WASP-19A, for those with compatible parallax ($\theta$) and proper motions (PM). Only one object was found to have parallax and proper motion values in both the RA ($\alpha$) and Dec ($\delta$) directions that are compatible with those of the target to within 5$\sigma$. This object ({\it Gaia} DR3 ID = 5411725145916602496), which is 8.8 magnitudes fainter in the $G$-band, has a parallax that differs from that of WASP-19A by $1.6\,\sigma$ and $\mathrm{PM_{\alpha}}$ and $\mathrm{PM_{\delta}}$ values  differing by just $0.1\,\sigma$ and $0.6\,\sigma$ respectively. This star is found in the southernmost part of the south-east region of WASP-19A, at a separation of $68^{\prime\prime}$. The magnitude and BP-RP colour of the companion suggests that this star is a mid M-dwarf, with a best matching spectral type for the magnitude of M6V and for the colour M2V, using the updated table\footnote{\url{https://www.pas.rochester.edu/~emamajek/EEM_dwarf_UBVIJHK_colors_Teff.txt}} of~\cite{Pecaut_Mamajek_13}.

To estimate the probability of this putative companion arising from a chance matching of WASP-19 in parallax and proper motions, we searched a much larger area of sky in $Gaia$ DR3, still centered on WASP-19, but with a radius of $2^\circ$. This produced 77 objects (including the companion; hereafter: `WASP-19B') that match WASP-19 within 5$\sigma$ in all of $\theta$, $\mathrm{PM_{\alpha}}$, and $\mathrm{PM_{\delta}}$. This implies that there is an 8.6\% chance of finding a matching object within our initial search radius of $4^\prime$, and just a 0.7\% chance of finding an object as close as WASP-19B. Relaxing the 5$\sigma$ limit to 3$\sigma$ changes these probabilities to 2\% and 0.16\%, respectively. Furthermore, none of these additional matches, all of which are more than ten times further from WASP-19A than WASP-19B, demonstrate as good a match as WASP-19B. Mathematically, this is expressed as WASP-19B having the lowest value of $\sigma_\mathrm{tot} = \sigma_{\theta} + \sigma_{\mathrm{PM_{\alpha}}} + \sigma_{\mathrm{PM_{\delta}}}$, where for each parameter, $X$ (where $X$ stands for $\theta$, $\mathrm{PM}_{\alpha}$, $\mathrm{PM}_{\delta}$) $\sigma_X = \frac{|X_1 - X_2|}{\sqrt{\sigma^2_{X_1} + \sigma^2_{X_2}}}$, where $X_1 \pm \sigma_{X_1}$ and $X_2 \pm \sigma_{X_2}$ are the values and uncertainties of parameter $X$ for WASP-19 and its companion, respectively.
We recognise that the uncertainty on the parallax of WASP-19B is relatively large. Even if WASP-19B is likely to be a true, gravitationally bound, companion to WASP-19A, some doubt remains as to whether it truly is a bound companion to WASP-19A. 
We adopted the planet name WASP-19Ab because it orbits around the brighter primary star.
Details of WASP-19B are given in Table~\ref{tab:companion}. With Eq.~(9) of~\cite{Csizmadia2019}, we also determined that the effect of the companion star on the radial velocity of star A is negligible. Star B may be the cause of the presence of an eccentricity different from zero in the orbit on the transiting planet. While tidal interaction leads to circularization, the excitation of the orbit due to a distant companion might prevent it, as discussed in detail in Appendix~\ref{app:eccentricity_prove}. 

\begin{table}
\resizebox{0.53\textwidth}{!}{
\renewcommand{\arraystretch}{1.4}
\begin{tabular}{lcc}
\hline
\hline
Parameter & WASP-19A & WASP-19B \\
\hline
Gaia DR3 ID & 5411736896952029568 & 5411725145916602496 \\
Apparent separation (\arcsec) & 0.0 & $68.0347\pm0.0009$ \\
Gaia $G$-magnitude & $12.1012$ & $20.8859$ \\
Gaia $BP-RP$ colour & $0.953\pm0.001$  & $2.213\pm0.12$ \\
pmRA (mas yr$^{-1}$) & $-35.4571\pm0.0089$ & $-35.33\pm1.18$ \\
pmDec (mas yr$^{-1}$) & $17.378\pm0.009$ & $18.22\pm1.30$ \\
Parallax (mas) & $3.7516\pm0.0090$ & $1.92\pm1.16$ \\
Distance (pc)$*$ & $266.54\pm0.64$ & $487\substack{+564 \\ -184}$ \\
Projected physical separation (AU) & 0.0 & $19976\pm48$ \\
Spectral type$^\dagger$ & G8V & M2V - M6V \\
\hline
\end{tabular}
}
\caption[]{WASP-19A and its stellar companion
\newline * Assuming no extinction.
\newline $^\dagger$ Assuming that the companion is also a star on the main-sequence.}
\label{tab:companion}
\end{table}

\section{Observations}
\label{sec:data}

\subsection{Transits and occultations}
\label{subsec:TrOcc}
Mid-transit and occultation times used in this paper are listed in Tables~\ref{Tab:LiteratureTR_appendix} - \ref{Tab:LiteratureOCC} in Appendix \ref{app:TROCCdata}. We collected publically available data, including some amateur observations listed in the {\it TRansiting ExoplanetS and CAndidates} (TRESCA)\footnote{\url{http://var2.astro.cz/EN/tresca/index.php}} project database, part of the {\it Exoplanet Transit Database} (\citealt{Poddany2010}). In particular, we included amateur data that were selected by~\cite{Mancini2013} and other high-quality data showing a clear light curve without gaps. As detailed in the tables, we also re-analyzed some transits with the Transit and Light Curve Modeller (TLCM,~\citealt{Csizmadia2020}).\\
\indent The time unit of each mid-transit and mid-occultation time was carefully checked and eventually converted into $\BJDTDB$ (barycentric Julian date, in the barycentric dynamical time system). The difference between $\BJDUTC$ and $\BJDTDB$ depends on the number of leap seconds \citep{Eastman2010}. For our dataset, it is around one minute. If the time unit used in the reference paper was not clear, the main author was contacted and the issue was clarified; otherwise, the data were discarded and not reported here. We used the estimated mid-times for each primary transit and occultation to evaluate whether they exhibit deviations from a Keplerian orbit over a long baseline of observations (see  Sect.~\ref{subsec:code} for more details) as well as to constrain the sidereal period of the planet. 

\subsection{RVs}
\label{sec:data_RVs}
\begin{figure} 
    \centering
    \includegraphics[width=0.55\textwidth]{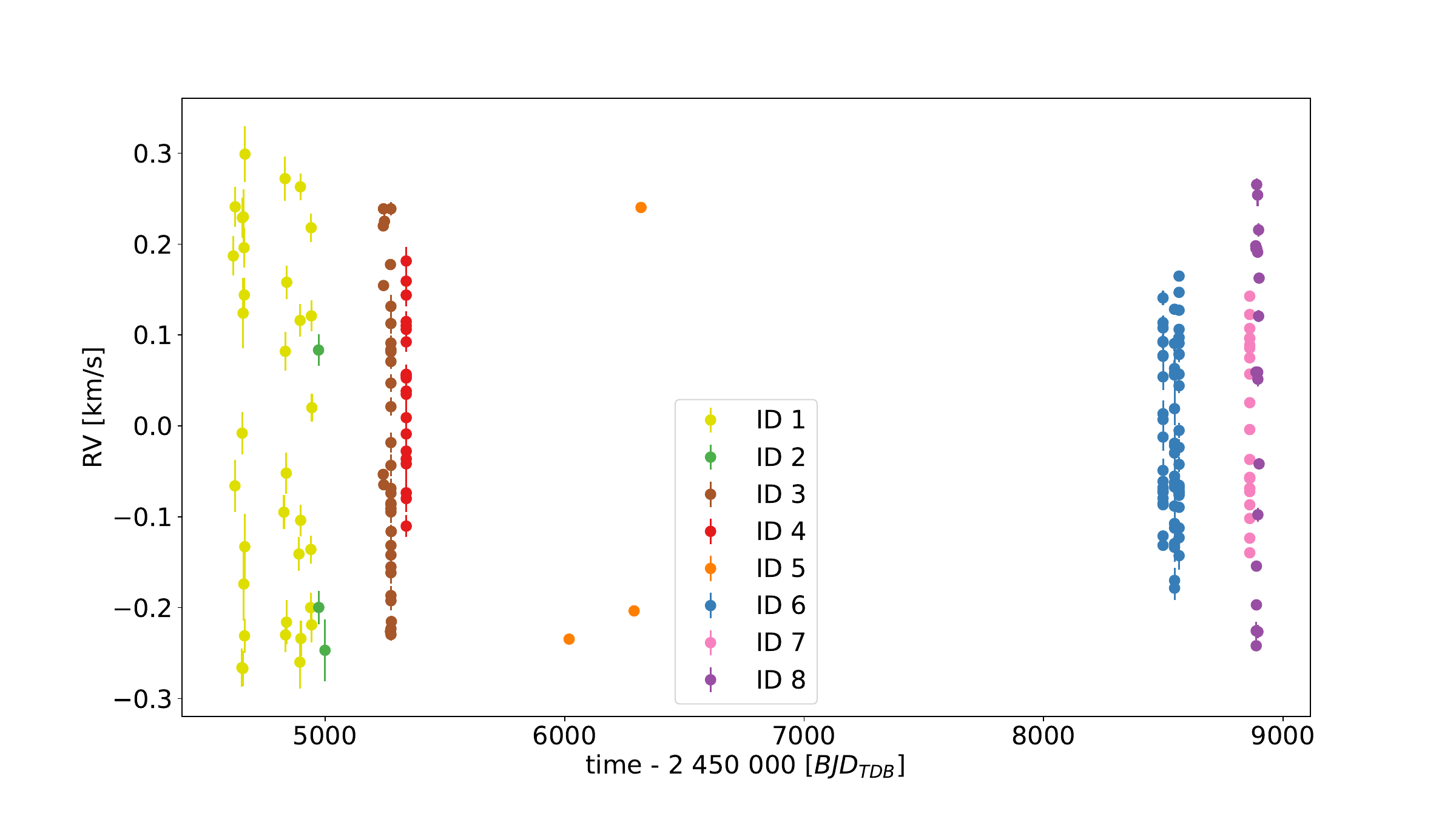}
    \caption{Radial velocity observations of WASP-19Ab, time vs radial velocities, with corresponding observational errorbars. The different data sets (see Sect.~\ref{sec:data}) are represented by different colours. 
    Between the datasets, there are significant offsets between different instruments and studies. They have been removed accordingly to the results in Table~\ref{tab:model_parameters}. 
    }
    \label{Fig:bjdtdb_RV}
\end{figure}

We collected all available radial velocity measurements from 2008 onward and we present new RV observations (see ID 8 below). 
Information on the RV observations are listed in Table~\ref{Tab:Datasets_RV} and the complete datasets are shown in Appendix~\ref{app:RVdata}.\\
\begin{table}[]
    \centering
    \begin{tabular}{c|c l l}
    \hline \hline
        ID & $N_\mathrm{obs}$ & Instrument & Reference  \\
        & (out of transit) & & \\
        \hline 
        1 & 34 (30) & Coralie &~\cite{Hebb2010} \\
        2 & 3 (2) & Coralie &~\cite{Hellier2011} \\ 
        3 & 36 (21) & HARPS &~\cite{Hellier2011} \\
        4 & 20 (6) & PFS &~\cite{Albrecht2012} \\
        5 & 3 (3) & HIRES &~\cite{Knutson2014} \\
        6, 7 & 88 (16, 6) & ESPRESSO &~\cite{Sedaghati2021} \\
        8 & 19 (15) & HARPS & This work \\
        \hline
    \end{tabular}
    \caption{Radial velocity datasets used in this work. \\
    RV datasets are described with ID number, number of points, instrument and reference for the publication of the corresponding data and results. In brackets, one can find the number of out-of-transit as those are the only RV datapoints that are used in the fit.}
    \label{Tab:Datasets_RV}
\end{table}
Here, we list the main details on the datasets, with the labels corresponding to the  discussion in Appendix~\ref{app:RVdata}: 
\begin{enumerate}
\item [ID 1]~\cite{Hebb2010} (discovery paper) reported 34 RV measurements obtained with the CORALIE spectrograph on the 1.2 m Euler telescope between 2008 and 2009. To the best of our knowledge and after contacting the authors of the paper, the dataset reported in the paper is in $\mathrm{BJD_{UTC}}$. We present the data transformed to $\BJDTDB$;\\
\item[ID 2]~\cite{Hellier2011}  reported 3  RV points from the CORALIE spectrograph on the Swiss Euler 1.2 m telescope in 2009. The data were reported in $\BJDUTC$ in~\cite{Hellier2011}, so they are transformed here into $\BJDTDB$; \\
\item[ID 3]~\cite{Hellier2011} reported 36 points during and around more than one transit in 2010 using the HARPS spectrograph on the ESO 3.6 m at La Silla, taken in order to measure the Rossiter-Mclaughlin effect. As well as for ID 2, the data were reported in $\BJDUTC$ in the original paper, so they have been transformed here into $\BJDTDB$; \\
\item [ID 4]~\cite{Albrecht2012} reported 20 data points taken in 2010 as measurements of the Rossiter-McLaughlin effect with the Magellan II (Clay) 6.5 m telescope and the Planet Finder Spectrograph (PFS); \\
\item [ID 5]~\cite{Knutson2014} reported 3 RV measurements obtained with the High Resolution Echelle Spectrometer (HIRES) on the 10 m Keck I telescope between 2008 and 2013; \\
\item [IDs 6,7]~\cite{Sedaghati2021} reported 88 RV points taken with VLT/ESPRESSO between 2019 and 2020, used to derive the Rossiter-McLaughlin effect and characterize the atmosphere of the planet. The two different IDs are given because of some setup changes in ESPRESSO, as described by~\cite{Sedaghati2021}, which lead to the introduction of a small offset. Moreover, as clarified with the first author of that paper, ID 6 was reported in $\BJDUTC$ in the original paper, while ID 7 in $\BJDTDB$, therefore the first one is transformed; \\
\item [ID 8] reported 19 RV points taken by us using the High Accuracy Radial velocity Planet Searcher (HARPS,~\citealt{Mayor2003}) at the ESO La Silla 3.6m telescope under the programme 0104.C-0849 (PI: Sz. Csizmadia). The observations were performed between 6 and 21 February 2020. We tried to obtain 1-3 points per night and we planned the observations to have a good phase coverage. The exposure time was 1200 seconds while the spectral resolution as 115,000. The {\it objAB} observing mode of HARPS was used, covering the 378-691 nm wavelength range. The measured signal-to-noise ratio (S/N) varied between 10.2 and 21.6 depending on weather and seeing conditions. The data were reduced by the standard HARPS pipeline available at La Silla Observatory and the results with their corresponding uncertainties are reported in Tables~\ref{Tab:dataRV3} and \ref{Tab:dataRV4}.
\end{enumerate}
\indent All RV observations are shown in Fig.~\ref{Fig:bjdtdb_RV}. As described in Sect.~\ref{subsec:code}, the datasets have different offsets \{which are subtracted in the figure. They are fitted in our model by including the term $V_\mathrm{{{i,instr}}}$ in Eq.~(\ref{eq:rv_model}). As for the transit and occultation mid-times, the time unit of the observation was transformed and uniformized to $\BJDTDB$ when necessary.\\

\section{Data analysis}
\label{sec:data_analysis}

\subsection{Long-term effects in RVs}
\label{subsec:long_term}


In this section, we study the long-term effects affecting the RV curve. If present, they must be identified and subtracted before modeling and estimating the rate of periastron precession in Sect.~\ref{subsec:code}. \\

\begin{figure}[h]
    \centering
    \includegraphics[width=0.53\textwidth]{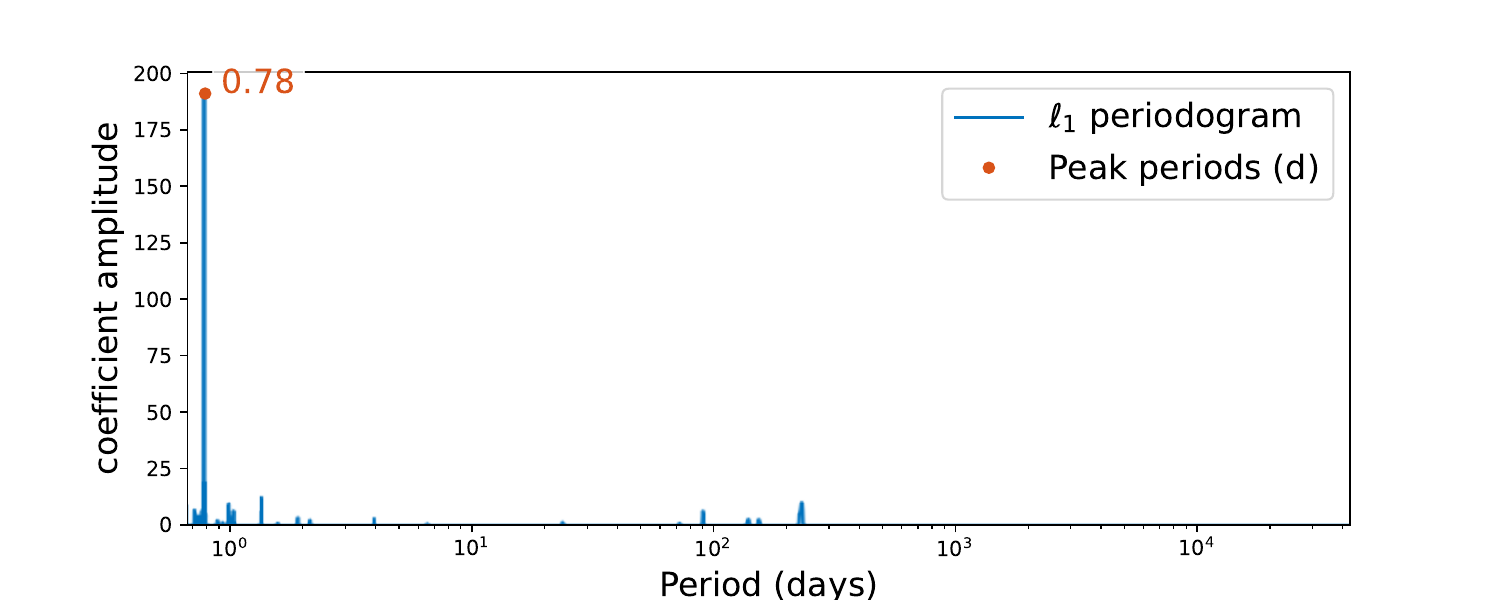}
    \includegraphics[width=0.53\textwidth]{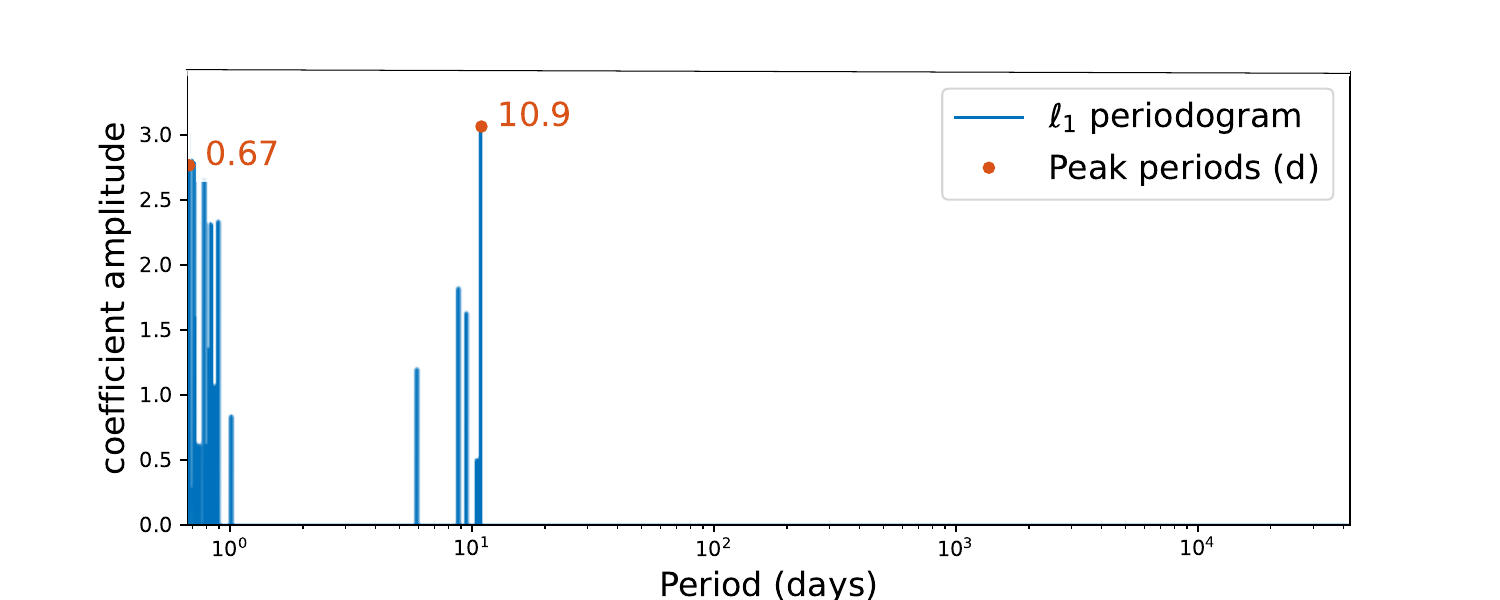}
    \caption { $l_1$ periodogram of all RV data outside the transit (top). There is only one significant peak with false alarm probability $\simeq 10^{-31}$.\ The periodogram of the residuals after the subtraction of the contribution of planet b (bottom). All residual peaks have false alarm probabilities of >20\%.}
    \label{fig:L1_periodogram}
\end{figure}

\indent First, we applied the $l_1$ periodogram (\citealt{Hara2017}) to the archive and newly acquired RV data (ID 8, see Fig.~\ref{fig:L1_periodogram}) to search for further companions in the system, 
in addition to the confirmed planet in the system. This method was developed specifically to find companions in RV datasets and can be used in a similar way to a Lomb–Scargle periodogram.  
To verify the significance of the peaks, we followed the re-sampling approach suggested by~\cite{Hara2017}: we randomly removed 10–20\% of the data and re-computed the periodogram. Some peaks did not appear after the re-sampling, therefore, we did not consider them real. 
After the re-sampling, we identified only one peak at 0.78 d, as shown in the top plot of Fig.~\ref{fig:L1_periodogram}. It was retrieved in correspondence with the orbital period of WASP-19Ab, with a false alarm probability (FAP) of $\mathrm{log_{10}(FAP)}$=-30.9 (FAP $\simeq 10^{-31}$). The analysis shows no evidence of any additional periodic signal in the RV data.\\
\indent We also re-computed the periodogram on the RV residuals after the subtraction of the signal of the first planet (see bottom plot in Fig.~\ref{fig:L1_periodogram}). After a re-sampling of the data, only a few small peaks are left. However, their FAP is $\geq$0.2, meaning that any peaks are due to artefacts.

\begin{figure}[h]
\centering
\includegraphics[width=0.45\textwidth]{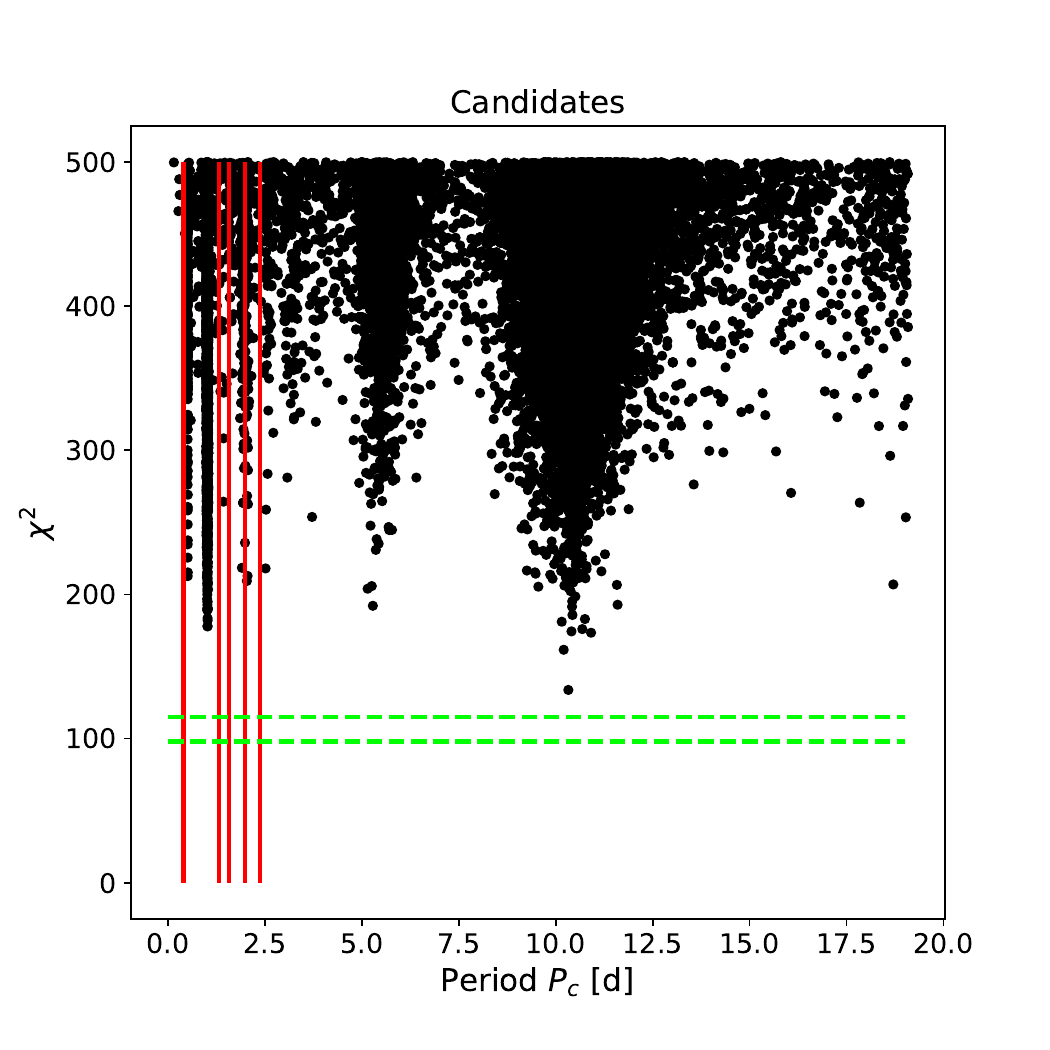}   \includegraphics[width=0.45\textwidth]{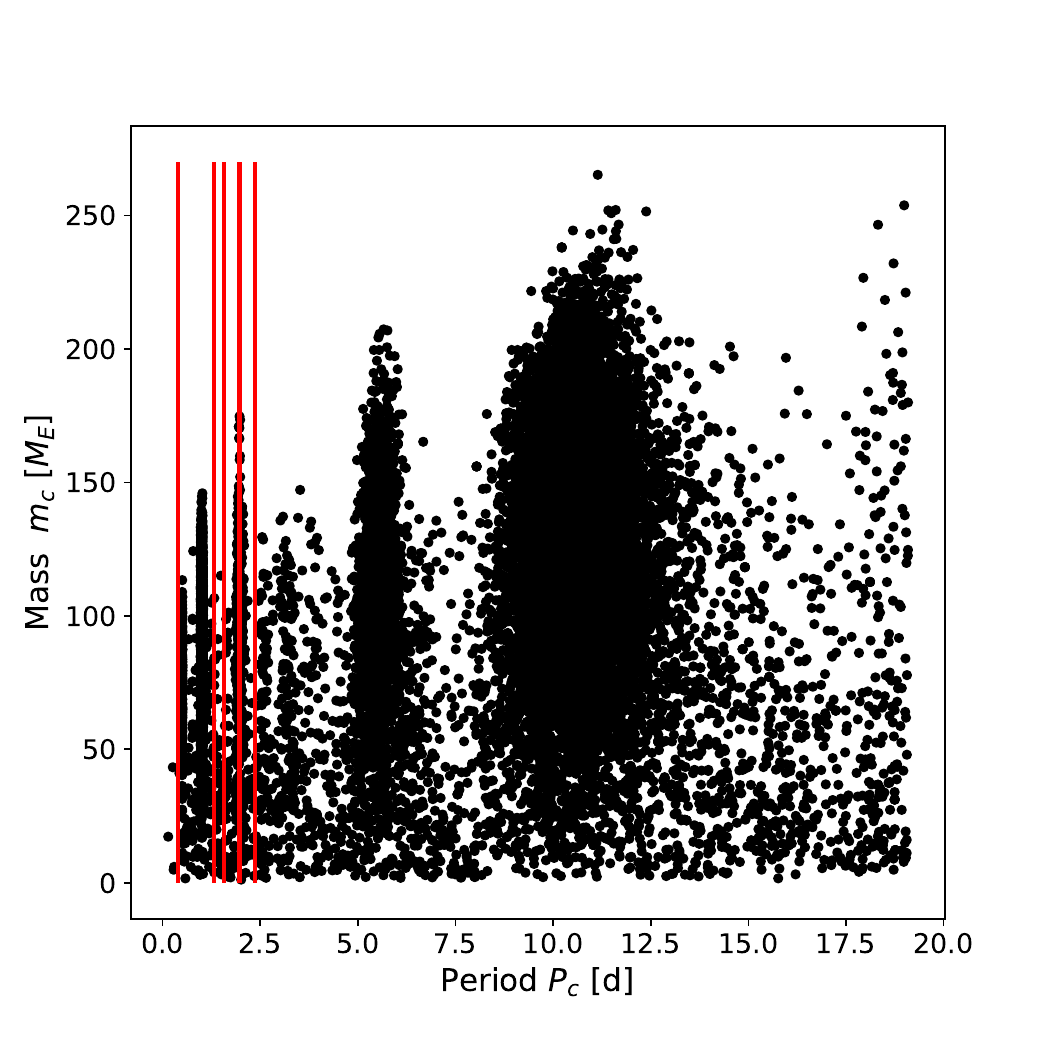}
    \caption{Properties of third bodies excluded by RV data. The plot shows the results for the simulated scenarios including a putative companion planet c of different masses at different periods. Each black dot represents one of the around 20\,000 selected scenarios (see the text for more details).\\
    {\it Top plot}: $\chi^2$ value vs the orbital period of the perturber. 
    The red lines correspond to the period of the stable orbits predicted by~\cite{CortesZuleta2020}. The two horizontal dashed lines represent the $\chi^2$ required to have weak (upper line) or strong (lower line) evidence of a second planet, as discussed in the text. \\
    {\it Bottom plot}: Mass and orbital period of the selected candidates. 
    }
   \label{fig:good_candidates}
\end{figure}

\noindent We estimated the sensitivity of our search for perturbing bodies in the data by performing a sensitivity study based on the transit timing variations. We calculated the maximum mass of a perturber in the system as a function of the maximum amplitude of the TTVs 
(see Eq.~(2) in~\citealt{HolmanMurray2005}). 
Our analysis shows that the maximum allowed mass of the perturber, $M_\mathrm{c}$, as a function of its orbital period, $P_\mathrm{c}$, assuming different values of the orbital eccentricity (e=0.0, 0.2, 0.4,0.6, and 0.8). The analysis shows that the region above 4000 $M_{\Jupiter}$ and period below 20 days is excluded by the TTV observations.\\
\indent We also performed a sensitivity study based on the RVs. The maximum amplitude of the RV residuals after the subtraction of the signal of the orbit of planet b (with \citealt{Sedaghati2021}'s parameters) is around 52 m s$^{-1}$. Through Kepler's third law, we estimated the maximum mass of the putative perturber. With an orbital period equal to 12 yrs (the total length of the RV dataset), we obtained $M_\mathrm{c} \simeq$ 3.9 $M_\mathrm{\Jupiter}$ and an orbital distance of $\simeq$ 5 AU.
We fitted the data with an additional Keplerian signal (beyond the signal of planet b) mimicking a hypothetical candidate for planet c. We injected a candidate and repeated the experiment two billion times with different random eccentric orbital parameters and fitted the new HARPS RV data. We varied the candidate parameters as follows: $\esin$ and $\ecos$ between -1.0 and 1.0, the RV semi-amplitude $K_\mathrm{c}$ between 0.1 and 100 m s$^{-1}$ (to widely investigate the amplitude of the RV residuals with errorbars) and the orbital period between 0.1 and the length of the dataset ($\simeq$19 d for the first attempt with the only HARPS dataset and $\simeq$ 12 yr for the second attempt with the whole dataset). Moreover, from Kepler's third law, we set the following condition to ensure that the orbit would not collide with the star:
\begin{equation}
    P_\mathrm{c} \geq \frac{2 \pi \Rs^{3/2}}{\sqrt{G\Ms}}, 
    \label{eq:constrain_on_P}
\end{equation}
where $G$ is the gravitational constant, while $\Rs$ and $\Ms$ are the radius and mass of the host star, respectively. 
With these constraints and varying the orbital parameters in the aforementioned ranges, the candidate mass range  is found to be between $\simeq$ 1.4 and $\simeq$ 400 M$_\Earth$. \\
\indent The results of our simulation (see Fig.~\ref{fig:good_candidates}) suggest scenarios that satisfy the condition $\chi^2 \leq 500$.
The top plot shows the $\chi^2$ of the model when including the proposed perturber, corresponding to five more free parameters: transit epoch, period, RV semi-amplitude, $\esin$ and  $\ecos$ (plotted as a function of the orbital period). 
Many of these candidates are discarded because the corresponding $\chi^2$ is too high or the amplitude of the perturbing body on the RV curve would be over a dozen m s$^{-1}$, so they would definitely be visible in the RV data and in the periodogram. The red lines underline the periods corresponding with the stable orbital regions present in the stability map by~\cite{CortesZuleta2020}. Their analysis allowed for the presence of a second planet either in 1:2, 5:3, 2:1, 5:2, or 3:1 resonance with planet b, corresponding to orbital periods of 0.39 d, 1.31 d, 1.58 d, 1.97 d, or 2.37, respectively. 
Some of these periods correspond with regions in the figure that are densely populated by our best candidates. 
The bottom plot shows the mass of the candidate plotted against its orbital period. Again, we highlight the stable orbit by~\cite{CortesZuleta2020} with red dashed lines.\\ 
\indent The expected $\chi^2$ would be around the number of degrees of freedom (19 total datapoints for dataset 8, of which 15 were fitted as out-of-transit points and 11 free parameters), while the lowest $\chi^2$ we obtained during the simulations is $\simeq$ 134 (clearly visible in the top plot of Fig.~\ref{fig:good_candidates} at $\simeq$10 d orbital period).
We fit the HARPS data with a one-planet eccentric orbit model to compare it to the scenarios with a perturbed. Without adding any RV jitter, the $\chi^2$ of this model is $\simeq$ 135. Since the two-planet model has four more free parameters, by applying the Bayesian information criterion (BIC), we found no evidence of the presence of a perturber in the system with a period within the length of the dataset with this approach. For weak (respectively, strong) evidence (see \citealt{Kass1995} for more details) that the two-planet model is preferred, a $\chi^2$ value below 115 (respectively, 98) is needed.\\ 
\indent We also injected a candidate in the whole RV dataset (IDs from 1 to 8), but we obtained an even higher $\chi^2$, resulting in an even more unlikely presence of a perturbed. Our results exclude it up to an RV amplitude of $\simeq$1 m s$^{-1}$.\\ 
\indent This simulation confirms the conclusion of~\cite{Knutson2014} when looking for the acceleration of the system due to possible long-period companions. The cited paper does not find an acceleration that differs from zero by more than 3$\sigma$. 
In this way, we extended the orbital period range of investigation in~\cite{Knutson2014} from between 0 and 1702 d up to $\simeq$4200 d.\\

\noindent Since we did not find any additional companion and there is no evidence for long-term trends in the data, we proceeded with our analysis by introducing periastron precession in the fit. 

\subsection{Rate of apsidal motion}
\label{subsec:code}
In this section, we describe the code used for the data analysis (by allowing for non-zero periastron precession) with the fitted parameters and the results.\\
\indent We used the approach and $\texttt{idl}$ code developed by~\cite{Csizmadia2019} to analyze the RV data in parallel with mid-transit and occultation times. Since then, the code has been modified and updated. Now the Genetic Algorithm minimization is followed by a differential-evolution Markov chain Monte Carlo (DE-MCMC,~\citealt{TerBraak2006}) algorithm to explore the parameter space. The values of the fitted parameters (see Table~\ref{tab:model_parameters} for the list of such parameters) obtained by the Genetic Algorithm are used as a starting point for the DE-MCMC analysis to derive the posterior probability distributions. The Genetic Algorithm is initialized with a population set made of 200 individuals. In the DE-MCMC we run 90 chains. The first 6$\cdot$10$^4$ steps of each chain are discarded as part of the burn-in phase. They are followed by 10$^4$ more DE-MCMC steps, which are used for the posterior distributions. The convergence is checked by the Gelman - Rubin test applied on the steps that follow the burn-in phase. At the end of the run, all parameters resulted in having R<1.2 \citep{GelmanRubin1992}. \\
\indent The priors that are used on the fitted parameters are set from the results in the literature (mainly from \citealt{Hellier2011}) and noted in Table~\ref{tab:model_parameters} along with the results. The prior on the eccentricity is set as the weighted mean and standard deviation of the results in the literature \citep{Hebb2010,Anderson2010,Anderson2013,Hellier2011,Mandell2013,Lendl2013,Zhou2014,Knutson2014}. From this, the priors on $\esin$ and $\ecos$ are calculated as a combination of the two parameters. The contribution on the $\chi^2$ of the priors was calculated the same way as in \cite{Csizmadia2019}.\\

\noindent We modeled the radial velocity of the host star with a Keplerian motion but with a time-variable $\omega$:
\begin{equation}
   V_\mathrm{rad,i,j} = V_\mathrm{\gamma} + V_\mathrm{i,instr} + K \cdot ( e \cdot \cos \omega (t_\mathrm{j}) + \cos (v_\mathrm{j} + \omega(t_\mathrm{j})) ) + \delta V(t_\mathrm{j}),
\label{eq:rv_model}
\end{equation}
with a secular advance of the longitude of periastron, $\dot{\omega}$ = $\mathrm{d\omega/dt,}$ caused by the tidal and rotational potential (as expressed in Eq.~(\ref{eq:domega/dt})):
\begin{equation}
   \omega(t_\mathrm{j}) = \omega_0 + \dot{\omega} (t_\mathrm{j}-t_0),
\label{eq:omega_model}
\end{equation}
where $i$ is the index of the instrument used and $j$ is the index of the time when the observation was taken; $V_\mathrm{\gamma}$ is the systematic velocity of the barycenter of WASP-19Ab system relative to the barycenter of the Solar System. Different datasets may have different instrumental zero-point offsets, $V_\mathrm{i,instr}$. Finally, 
$K$ is the radial velocity half-amplitude:
\begin{equation}
    K = \frac{2 \pi a_\mathrm{\star} \sin i}{P_\mathrm{a} \sqrt{1-e^2}}.
    \label{eq:RVsemiamplitude}
\end{equation}
In this expression, $a_\mathrm{\star}$ is the semi-major axis of the star around the common center of mass and $i$ is the inclination between the orbital plane and the plane of the sky ( perpendicular to the line of sight). \\
\begin{figure}
    \centering
    \includegraphics[width=0.5\textwidth]{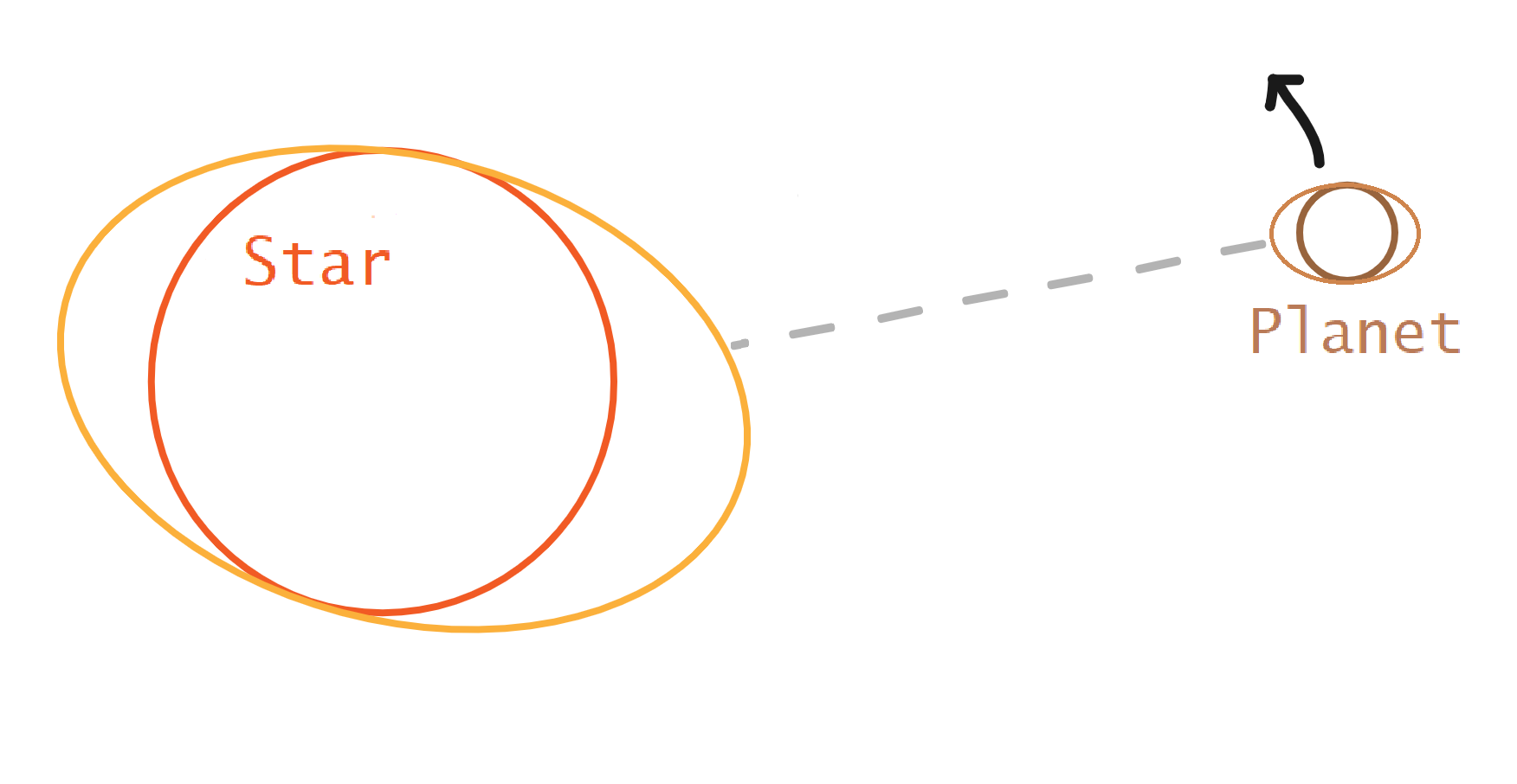}
    \caption{Star-planet system geometry when tidal interactions occur. The spherical shapes that star and the planet would have in the absence of tidal interactions are drawn in red and dark brown.  The tidal bulges are drawn in orange and light brown, respectively. They arise on the stellar and planetary surfaces (respectively in orange and light brown) and rotate with the star itself and are elongated in the direction of the other body. The tidal lags are also drawn. 
    The image is not to scale.}
    \label{fig:system_tidal_int}
\end{figure}

\noindent Here, $\delta V$ is the non-orbital apparent contribution that arises from the rotation of the star around its own axis. 
The stellar and planetary shapes are distorted ellipsoids with the longest axis in the direction of the radius vector connecting them, as shown in Fig.~\ref{fig:system_tidal_int}. Their apparent area varies continuously during an orbital revolution, therefore the emitted light changes (i.e., the ellipticity effect). Moreover, in close binary systems, part of the radiation of each of the two components falls on the surface of the companion. It is absorbed and re-emitted (or scattered) in all directions, 
varying during the orbital phase (i.e., the reflection effect).\\
\indent These two photometric effects in the RV data due to distortion are described by~\cite{Kopal1959} as a non-orbital contribution to the RV curve:
\begin{align}
\begin{split}
    \delta V = - \frac{3}{2} K \hspace{0.1cm} \sqrt{1-e^2} \hspace{0.1cm} \frac{\Ms+\mpl}{\Ms} \hspace{0.1cm}  \frac{\Porb}{P_\mathrm{rot,\star}} \hspace{0.05cm} \left(\frac{\Rs}{a}\right)^4   \cdot \\ \hspace{1cm} \cdot \left(1 + \frac{k_{2,\star}}{2}\right) \sin i \cdot \sin \hspace{0.05cm} (2 \hspace{0.05cm}(v + \omega)) f_2,
    \label{eq:app_eff}
\end{split}
\end{align}
where the Love number of the star is assumed from theoretical calculations (we take it from \citealt{Claret2004}). Then, $\Prots$ is the rotational period of the star around its axis and $f_2$ takes into account the limb darkening. For more details on $f_2$ see~\cite{Kopal1959} and~\cite{Csizmadia2019}.\\
\indent~\cite{Arras2012} already includes such calculations for precise RV analysis of WASP-18A, but neglects the effect of stellar rotation. In other words, a synchronous rotation of the star is assumed. 
However, for WASP-19A this is not the case. It is known to be active, showing a rotational modulation with $\Prots$ = 10.5 d (\citealt{Hebb2010}). Therefore, the term $\Porb$/ $\Prots$ is not expected to be 1, but $\simeq$ 0.79 d/10.5 d $\simeq$ 0.08, introducing a factor of about 12 difference in the amplitude of such effect calculated by~\cite{Arras2012} and~\cite{Kopal1959}.
The amplitude of this contribution is around 0.19 m s$^{-1}$, so it is negligible for WASP-19A if compared to the typical observational errorbars. The period of this effect is linked to the sine term of the true anomaly in Eq.~(\ref{eq:app_eff}) and, therefore, it is half of the orbital period. \\

\noindent Moreover, we added jitter terms for each spectrograph to the RV model (see, e.g., \citealt{Baluev2009} for a discussion of instrumental jitters). The jitter is composed of an instrumental and a stellar contribution. It is not possible to separate them. As a starting point, if present, we used the values proposed by the corresponding literature. In particular, 13 m s$^{-1}$ for~\cite{Hebb2010}'s data (ID 1, as in \citealt{Anderson2010}), 14.1 m s$^{-1}$ and 6.9 m s$^{-1}$ for Coralie (ID 2) and HARPS (ID 3)~\cite{Hellier2011}'s data, respectively, 20 m s$^{-1}$ for~\cite{Albrecht2012}'s (ID 4) and 17.8 m s$^{-1}$ for~\cite{Knutson2014}'s (ID 5) RVs.  
Then, we refined all the jitter values by minimizing the $\chi^2$ as:
\begin{equation}
    \chi^2_\mathrm{jit, instr} = \sum_\mathrm{i=1}^{n_\mathrm{data}} \frac{(O_\mathrm{i}-C_\mathrm{i})^2}{\sigma_\mathrm{i}^2 + \sigma_{\mathrm{jit_{instr}}}^2} = n_\mathrm{data} - 1,
    \label{eq:sigma_jitt}
\end{equation}
where $O_\mathrm{i}$ and $C_\mathrm{i}$ are the observed points (see Tables~\ref{Tab:dataRV1} and \ref{Tab:dataRV2}) and the (calculated) expected values from the model, respectively; $\sigma_\mathrm{i}$ represents the observed errors corresponding to $O_\mathrm{i}$ and $n_\mathrm{data}$ is the number of datapoints in the subset. This procedure was iteratively repeated until the condition of Eq.~(\ref{eq:sigma_jitt}) was fulfilled. Due to the highly inhomogeneous set of data, this procedure allows for every RV dataset to have a weight in the fit proportional to the number of its data points. \\
\indent We also added a jitter value for transits and one for occultations to account for inhomogeneities in the dataset and for a perturber with an orbital period longer than 12 years, as described in Secion~\ref{subsec:long_term}. The fitted values for the jitter terms can be found in Table~\ref{tab:model_parameters}, along with all the fitted and derived parameters in our model.\\
\\ 

\noindent The orbital eccentricity is a key issue for Love number studies: in circular orbits, apsidal motion is not observable. In addition, the argument of periastron is hard to measure if eccentricity is not well known. When the eccentricity is small, then its uncertainty range is often large and the investigators find that the orbit is compatible with a circular one. However, such results in the literature do not necessarily imply that the orbit is circular, despite many investigators setting a zero eccentricity (and not propagating the error bars of eccentricity to the corresponding result). It only means that an upper limit exists for eccentricity.\\
\indent Moreover, a certain dataset may indicate the presence of an eccentric orbit with an accurate result. However, the sample size is not big enough or not aptly distributed to get the uncertainty ranges accurate enough. The results are imprecise and the real value of eccentricity is vague.\\
\indent Although we collected more RV and TTV data than any previous study of the WASP-19Ab system, the datasets come from various instruments with different precisions and, thus, are non-homogeneous. The value of eccentricity may be affected by the different data accuracy, as some of the data sets are diluted by less precise measurement points. Because of these reasons, any previous estimate of eccentricity can be accurate but imprecise. \\
\indent Applying a prior on the eccentricity in the fitting procedure can be a valid approach. The average and standard deviation of the previously reported eccentricities can better represent reality than any single determination from a limited dataset. Therefore, we calculated the weighted average of the previous eccentricity measurement and their weighted standard error using their uncertainty range as weights. Therefore, we performed a fit using this mean eccentricity as a prior with their weighted standard error as a Gaussian width. \\ 
\indent In addition, in the calculation of the $\chi^2$ of the fit, priors also on the stellar temperature, $\Teffs$, and Love number, $\Loves$, the mass ratio, $q$, the orbital inclination, $i$, the orbital over rotational period ratio, $\mathrm{F_\star}$, and the ratio, $\rp/\Rs$, were added as described in~\cite{Csizmadia2019}.\\


\begin{table*}[]
    \centering
    \begin{tabular}{l l l c l}
    \hline 
    \hline 
        Parameter & Symbol & Fit & Value & Priors\\
        \hline
        {\bf Assumed parameters} & & & &\\
        Stellar rotation & $V\sin(i)$ & & 4.63 km s$^{-1}$ $^{\dagger 1}$ \\
        Stellar mass & $M_{\star}$ &  & 0.970 M$_{\sun}$\\
        Planet-to-star radius ratio & $r_\mathrm{p}$/$R_{\star}$ & & 0.14248 \\
        Semi-major axis & $a$ & & 0.01655 AU \\
        First transit epoch & $T_0$ &  & 2~454~775.3372 $\BJDTDB$ \\
        \\
        {\bf Adjusted parameters}\\
        Jitters \hspace{3cm} ID 1 & jit$_1$ & RVs & 13.5 m s$^{-1}$  \\
        \hspace{3.93cm} ID 2 & jit$_2$ & RVs & 35.0 m s$^{-1}$  \\
         \hspace{3.93cm} ID 3 & jit$_3$ & RVs & 9.8 m s$^{-1}$ \\
         \hspace{3.93cm} ID 4 & jit$_4$ & RVs & 11.5 m s$^{-1}$ \\
         \hspace{3.93cm} ID 5 & jit$_5$ & RVs & 30.4 m s$^{-1}$  \\
         \hspace{3.93cm} ID 6 & jit$_{6}$ & RVs & 13.1 m s$^{-1}$  \\
         \hspace{3.93cm} ID 7 & jit$_{7}$ & RVs & 9.8 m s$^{-1}$ \\
        \hspace{3.93cm} ID 8 & jit$_{8}$ & RVs & 22.3 m s$^{-1}$  \\ 
         \hspace{3.93cm} Transits & jit$_\mathrm{tr}$ & Tr & 0.00005 d\\
         \hspace{3.93cm} Occultations & jit$_\mathrm{occ}$ & Occ & 0.0001 d\\
         
        {\bf Fitted parameters}\\
        Offset velocity of the system & $\mathrm{V_{\gamma}}$ & RVs & (20.7791 $\pm$ 0.0060) km s$^{-1}$ & $\mathcal{U}$ [-$\infty$,+$\infty$]\\
        RV semi-amplitude & K & RVs & (253.48 $\pm$ 3.87 ) m s$^{-1}$ & $\mathcal{U}$ [-$\infty$,+$\infty$]\\
        Offset instrumental velocity \\ 
        \hspace{3.93cm} ID 2 
        & V$\mathrm{_{instr,2}}$ & RVs & (0.0258 $\pm$ 0.0071)km s$^{-1}$ & $\mathcal{U}$ [-$\infty$,+$\infty$]\\
        \hspace{3.93cm} ID 3 & V$\mathrm{_{instr,3}}$ & RVs & (0.0221 $\pm$ 0.0074 ) km s$^{-1}$ & $\mathcal{U}$ [-$\infty$,+$\infty$]\\
        \hspace{3.93cm} ID 4 & V$\mathrm{_{instr,4}}$ & RVs & (-20.827 $\pm$ 0.016 ) km s$^{-1}$ & $\mathcal{U}$ [-$\infty$,+$\infty$]\\
        \hspace{3.93cm} ID 5 & V$\mathrm{_{instr,5}}$ & RVs & (-20.712 $\pm$ 0.014 ) km s$^{-1}$ & $\mathcal{U}$ [-$\infty$,+$\infty$]\\
        \hspace{3.93cm} ID 6 & V$\mathrm{_{instr,6}}$ & RVs & (-0.0919 $\pm$ 0.0092 ) km s$^{-1}$ & $\mathcal{U}$ [-$\infty$,+$\infty$]\\
        \hspace{3.93cm} ID 7 & V$\mathrm{_{instr,7}}$ & RVs & (-0.1149 $\pm$ 0.0139 ) km s$^{-1}$ & $\mathcal{U}$ [-$\infty$,+$\infty$]\\
        \hspace{3.93cm} ID 8 & V$\mathrm{_{instr,8}}$ & RVs & (0.0750 $\pm$ 0.0090) km s$^{-1}$ & $\mathcal{U}$ [-$\infty$,+$\infty$]\\
    
        Anomalistic period & $P_\mathrm{a}$ & RVs and Tr/Occ & (0.788951 $^{+0.000012}_{-0.000017}$) d & $\mathcal{U}$ [-$\infty$,+$\infty$]\\  
        & $\esin$ & RVs and Tr/Occ & 0.03920 $^{+0.00458}_{-0.00682}$ & $\mathcal{U}$ [-1,+1] \\
        & $\ecos$ & RVs and Tr/Occ & -0.00056 $^{+0.01385}_{-0.01717}$ & $\mathcal{U}$ [-1,+1]\\
        Periastron precession rate & $d\omega/dt$ & RVs and Tr/Occ & (0.065 $^{+0.007}_{-0.010}$)$^\circ$ d$^{-1}$ = \\
        & & & (233 $^{+25}_{-35}$)$^{\prime\prime}$ d$^{-1}$ & $\mathcal{U}$ [-$\infty$,+$\infty$] \\
        Stellar Love number & $k_{2,\star}$ & RVs & 0.0200 $\pm$ 0.0069 & $\mathcal{N}$ (0.02;0.004)\\
        Stellar temperature & $\Teffs$ & RVs & (5498.65$\pm$ 172.86)K &  $\mathcal{N}$ (5500 K;100 K) \\
        Planet-to-star mass ratio & $q$ = $m_\mathrm{p}$/$M_{\star}$ & RVs & 0.001150 $\pm$ 0.000057 & $\mathcal{N}$ (0.00115;3.3$\cdot$10$^{-5}$) \\ 
        Stellar radius over semi-major axis & $R_{\star}$/$a$ & RVs & 0.2785$
        \pm$ 0.0070 & $\mathcal{N}$ (0.2784;0.0040)\\
        Orbital-to-rotational stellar period & $F_{\star}$=$\Porb$/$\Prots$ & RVs & 0.0729 $\pm$ 0.0074 & $\mathcal{N}$ (0.0729;0.0043) \\
        Orbital inclination & $i$ & RVs and Tr/Occ & (79.40 $\pm$ 0.69)$^\circ$ & $\mathcal{N}$ (79.4;0.4)\\
        
        \\
    {\bf Derived parameters} \\
    Eccentricity & e & & 0.00172 $^{+0.00035}_{-0.00033}$ & \\
    Periastron angle & $\mathrm{\omega_0}$ & & (89.71$^{+24.68}_{-19.12}$ )$^{\circ}$ \\ 
    Sidereal period & $P_s$ & & (0.788839267$^{+0.000000018}_{-0.000000019}$ ) d \\
    Planetary Love number & $\Lovep$ & & 0.20 $^{+0.02}_{-0.03}$\\
    \hline 
    \end{tabular}
    \caption{Parameters of the WASP-19Ab system.
    Assumed, adjusted -- according to Eq.~(\ref{eq:sigma_jitt}) -- fitted parameters and their priors, and derived parameters for our model which fits RVs and transit/occultation mid-times in parallel. The third column indicates if the parameter was fitted to model radial velocity ("RV") and/or mid-transit and mid-occultation times ("Tr/Occ"). The last column shows the results of our analysis. The offset related to ID 1 ($V_{instr,1}$) was not included because the first dataset was used to derive the relative velocity between the observer and the center of mass of the system ($V_{gamma}$).\\
    The solutions are calculated only from those steps of the DE-MCMC that fall within the distribution around the highest peak of the posterior distribution of $\dot{\omega}$ (see Fig.~(\ref{fig:wdot_period}), solution B).\\
    Errorbars are given with 1$\sigma$ confidence level.\\
    The reported values and errorbars of the sidereal period $P_s$, $\omega_0$ and $e$ are not calculated - as $P_s$=$P_a \cdot(1-P_a \cdot \dot{\omega}/ n)$, $\arctan(\esin/\ecos)$ and $(\esin)^2$ + $(\ecos)^2$ respectively - from the reported values of $P_a$, $\dot{\omega}$, $\esin$ and $\ecos$ but from the median of the distribution resulting from the combination of the posterior distributions of $\esin$ and $\ecos$.\\ 
    $^{\dagger 1}$~\citealt{Hellier2011}\\
    }
    \label{tab:model_parameters}
\end{table*}

\noindent Together with the RVs, we modeled mid-transit and mid-occultation times listed in Tables~\ref{Tab:LiteratureTR_appendix} - \ref{Tab:LiteratureOCC} mainly to refine the determination of the first transit epoch and the period. To include periastron precession in the fit of predicted mid-times, we followed the approach of~\cite{Gimenez1983} and~\cite{Csizmadia2019}. 

\subsection{Presence of apsidal motion}
\label{subsec:presenceofwdot}
\begin{figure}
    \centering    
    \includegraphics[width=0.5\textwidth]{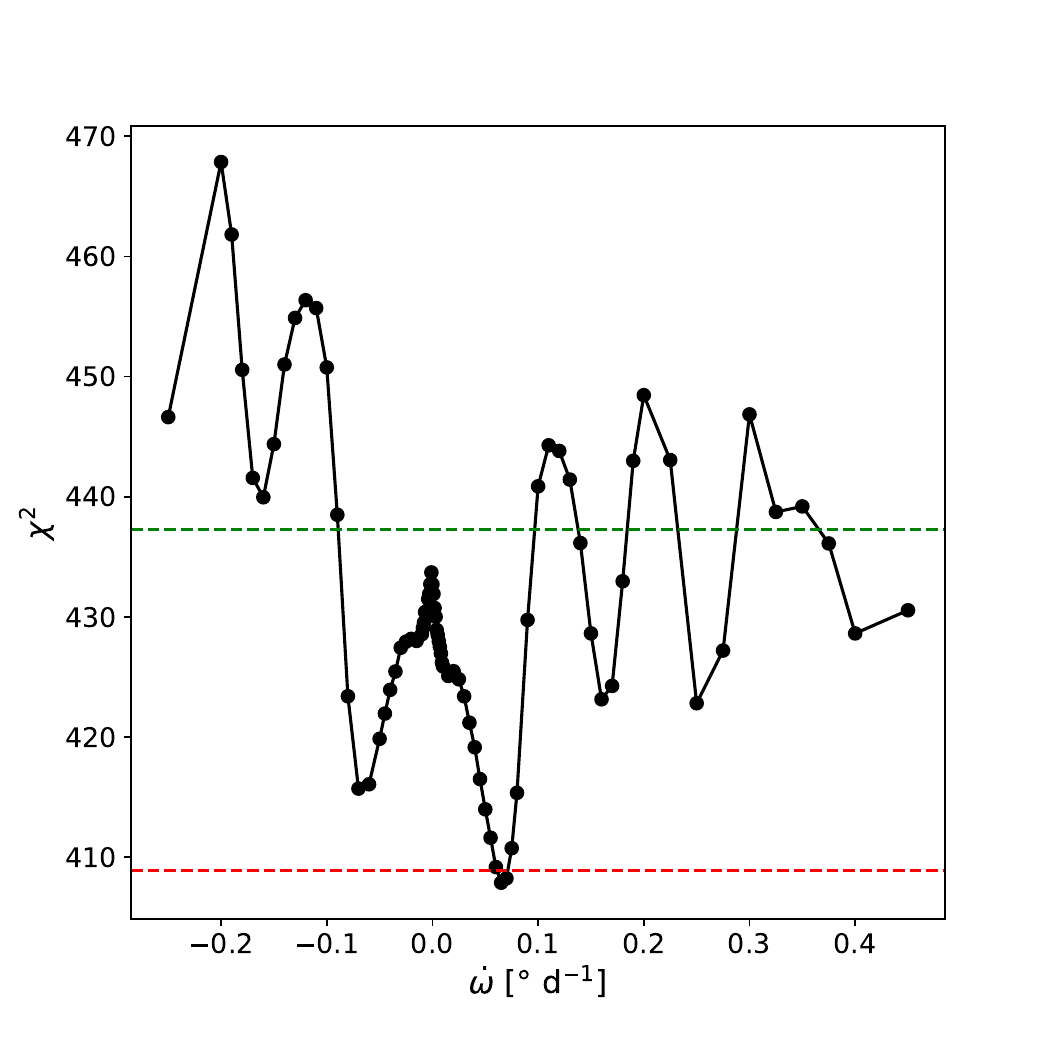}
    \caption{$\chi^2$ values of the fit when varying $\dot{\omega}$ in the model. The horizontal dashed line separates the statistically worse models (over the line) from the best models (below the dashed lines). The models below the dashed lines are statistically equivalent within 1$\sigma$ (68\% confidence level for the red line) and 2$\sigma$ (95\% confidence level for the green line). Note: the significantly larger $\chi^2$ at $\dot{\omega}$=0. The high amount of local minima is an indication of the complexity of the problem.}
    \label{fig:semi_grid}
\end{figure}

Because of the large parameter correlations between the mean motion and periastron precession rate (see~\citealt{Csizmadia2019}), as a first analysis, we calculated 80 different models with fixed values of $\dot{\omega}$, the so-called semi-grid analysis. This procedure also helps to get an initial range for the fitted value of $\dot{\omega}$. The limiting values were $\dot{\omega} = -0.25$ and $+0.45^\circ $d$^{-1}$. We also fixed the epoch to 2\,454\,775.3372 $\BJDTDB$, while all other parameters were free in the DE-MCMC analysis. The result is presented in Fig.~(\ref{fig:semi_grid}), showing the assumed value of the periastron precession rate and the corresponding $\chi^2$ of the fit. The figure clearly shows that the joint fit of the RV and TTV data is significantly worse if $\dot{\omega} = 0$, namely, no periastron precession is present. A small amount of periastron precession improves the quality of the fit significantly. The figure shows that small negative periastron precession rates also improve the fit but still result in a statistically worse fit than the small positive precession values. From that figure and the statistical analysis, we conclude that small positive periastron precession rates are needed to have a good fit of the joint RV and TTV data of WASP-19Ab, and we are able to give a lower and an upper limit of $0.06\lesssim\dot{\omega}\lesssim0.075^\circ $d$^{-1}$. The result is refined in the next section.

\section{Discussion of the results}
\label{sec:discussion}

\subsection{Apsidal motion rate}

In this section, we show and discuss the results of the analysis performed with the model described in Sect.~\ref{sec:data_analysis}. Contrarily to the semi-grid method (Sect.~\ref{subsec:presenceofwdot}), here we fit~$\dot{\omega}$ along with the other parameters.\\

\begin{figure}
    \centering \includegraphics[width=0.55\textwidth]{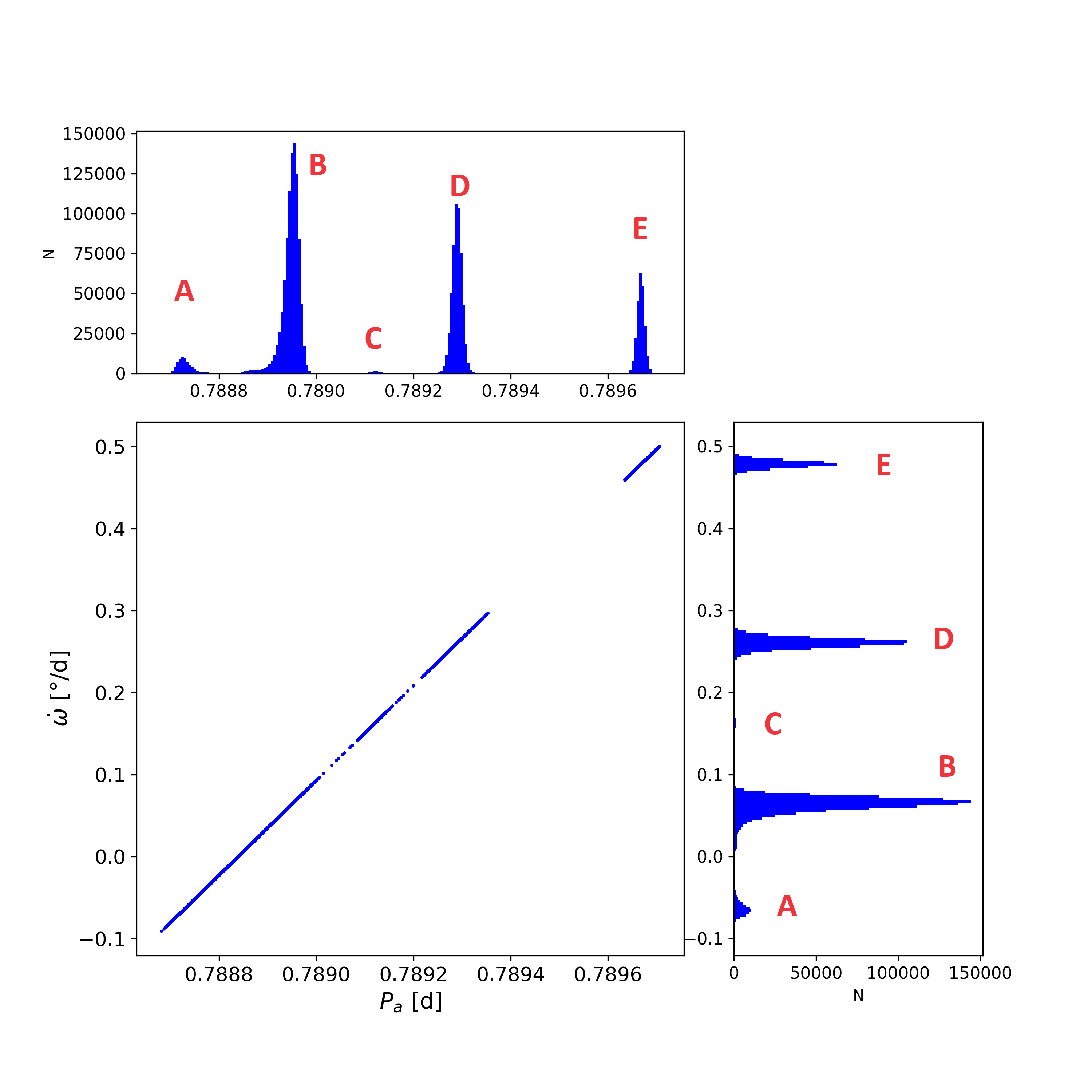}
    \caption{
    Posterior distributions and correlation between the anomalistic period and periastron precession rate where the chains converged. The middle plot shows the strong correlation between these two free parameters, as in Eq.~(\ref{eq:Ps_Pa}). On the sides, are the corresponding histograms of the posterior distributions. 
    As can be seen from the histograms on the sides, many chains converged to various positive and negative peaks, as predicted by the semi-grid method and reported in Table~\ref{tab:five_solutions}.
    }
    \label{fig:wdot_period}
\end{figure}

\noindent During the analysis, we noticed that different DE-MCMC chains converged to different values of $\dot{\omega}$. In particular, as it is shown in Fig.~\ref{fig:wdot_period} and in Table~\ref{tab:five_solutions}, during the run multiple chains explored various positive and negative values of $\dot{\omega}$. 
\noindent Figure~\ref{fig:wdot_period} clearly shows also the correlation between the anomalistic period and the periastron precession rate, as discussed in~\cite{Csizmadia2019}. It is due to the formula that links the anomalistic period $P_\mathrm{a}$ and the sidereal period $P_\mathrm{s}$:
\begin{equation}
    P_\mathrm{s} = P_\mathrm{a} (1-\dot{\omega}/n).
    \label{eq:Ps_Pa}
\end{equation}
This formula that links the two quantities gives a between period and periastron precession rate, but no unique solution. The added value of including transit and occultation mid-times lies in the fact that they constrain the sidereal period, and therefore a range of values for $P_\mathrm{a}$ and $\dot{\omega}$ (as shown in Fig.~\ref{fig:wdot_period}), while avoiding some other ranges.\\
\noindent A different chain convergence shows why it was important to run various chains. Only one chain would have converged to one of the possible values, excluding the other results or could have been trapped in a local minimum (the same local minima found by the semi-grid method approach; see Fig.~\ref{fig:semi_grid}). \\
\indent The highest peak of $\dot{\omega}$ (solution B) in the posterior distribution results around $\sim$+0.065$^\circ $d$^{-1}$ (summarized in Table~\ref{tab:five_solutions}). The positive solutions C, D, and E can be discarded because of their high $\chi^2$ values. Moreover, solution E would lead to a planetary Love number higher than 1.5, which is unphysical.
\begin{table}[]
    \centering
    \begin{tabular}{c c c}
    \hline \hline
        Solution & $\dot{\omega}$ [$^{\degree} d^{-1}$] & $\chi^2$ \\
        \hline
        A & -0.064 & 416 \\
        B & +0.065 & 408 \\
        C & +0.163 & 423 \\
        D & +0.261 & 424 \\
        E & +0.479 & 430 \\
        \hline
    \end{tabular}
    \caption{Median values of the five posterior distributions shown in Fig.~\ref{fig:wdot_period} and corresponding $\chi^2$ (calculated before introducing the contribution of the jitters to the uncertainty, as in Eq.~(\ref{eq:sigma_jitt})).
    }
    \label{tab:five_solutions}
\end{table}

\begin{figure}
    \centering
    \includegraphics[width=0.5\textwidth]{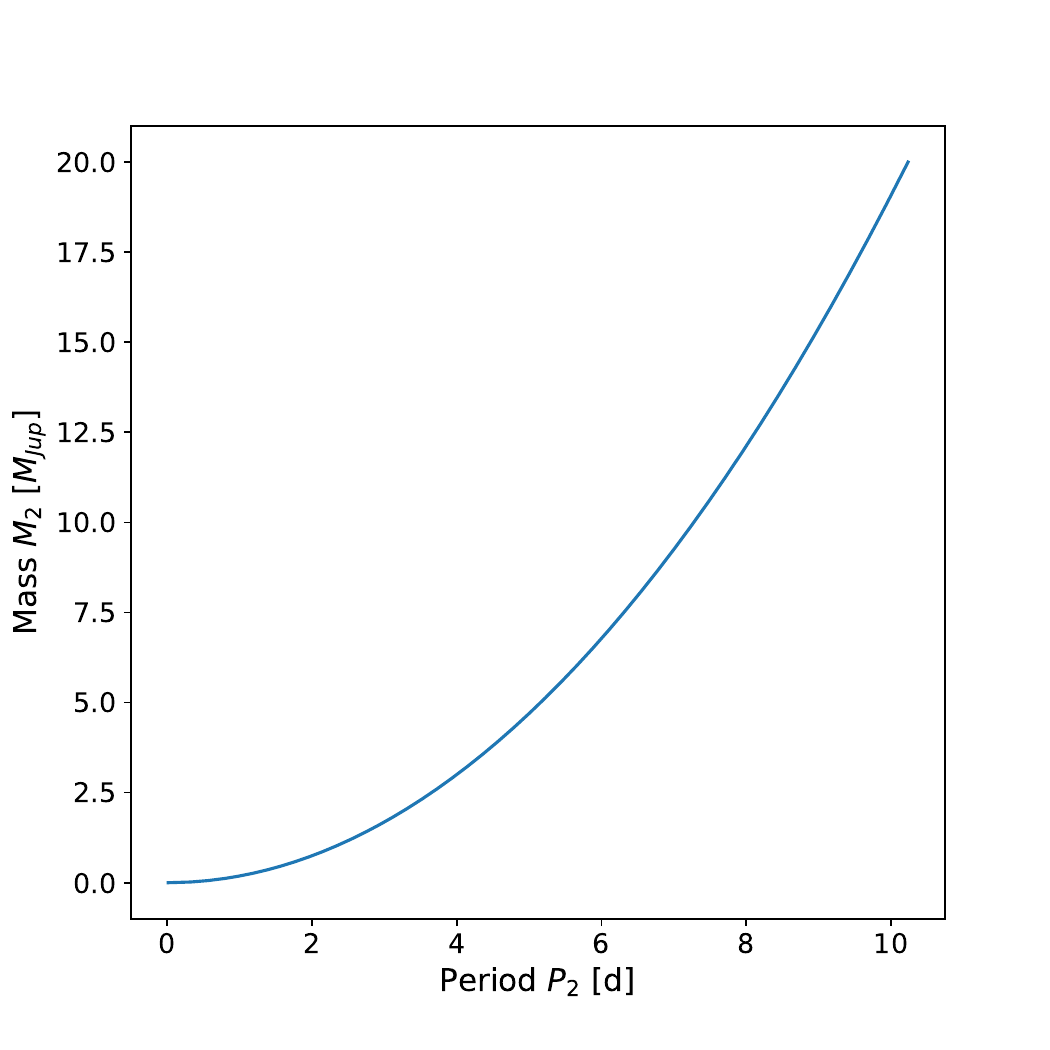}
    \caption{Mass and orbital period of a second body that would cause a negative periastron precession rate of $\sim$-0.0638$^\circ$d$^{-1}$ (corresponding to the negative solution of the DE-MCMC analysis) on planet b, as deduced from~\cite{Borkovits2011}. See text for more details.}
    \label{fig:negative_wdot}
\end{figure}

\begin{figure}[!]
    \centering
   \includegraphics[width=0.53\textwidth]{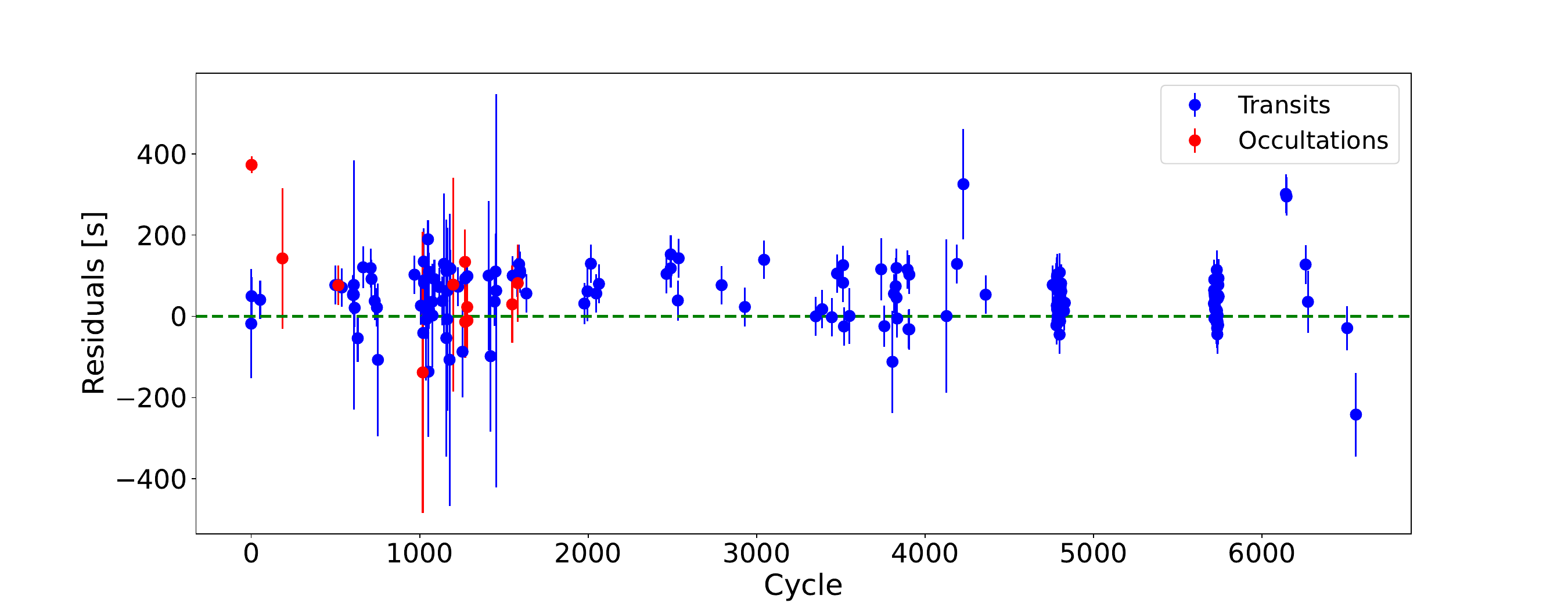}
    \caption{
    Residuals of the periastron precession fit for transits and occultation mid-times, calculated with the sidereal period. Transit mid-times are plotted as blue circles, occultation mid-times as red circles. The errorbars include the jitters, as in Table~\ref{tab:model_parameters}.\\
    }

    \label{fig:OC}
\end{figure}

\begin{figure}
    \centering
    \includegraphics[width=0.52\textwidth]{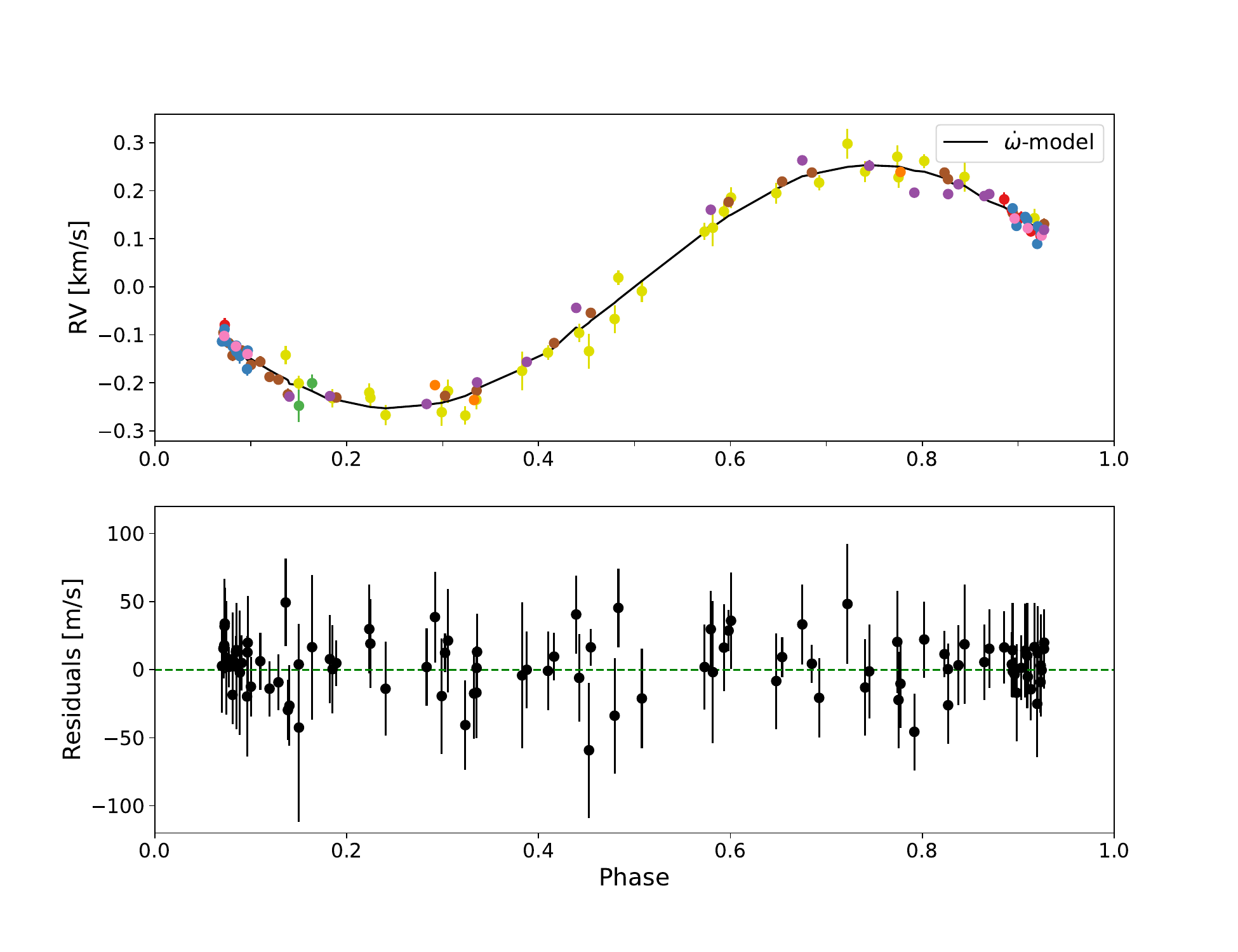}
    
    \caption{
    Data with errorbars, phase-folded with our fit including periastron precession, shown in the upper plot. The colours represent different datasets, as in Fig.~(\ref{Fig:bjdtdb_RV}) (see also Sect.~\ref{sec:data_RVs} for more details). \\
    Residuals of the periastron precession rate fit are shown in the lower plot. The errorbars include the jitters, as in Table~\ref{tab:model_parameters}.\\
    In both plots, the in-transit points are not shown as they were not used in the fit.
    }
    \label{fig:phase_folded_RV}
\end{figure}

\indent The negative periastron precession rate (solution A) would correspond either to a retrograde orbit (of the planet with respect to the rotation of the star) or to an unseen perturber. The retrograde orbit can be excluded from Rossiter-McLaughlin effect studies and TESS light curves \citep{Wong2020}. They confirm that the spin axis of the star is almost perpendicular to the orbital plane.\\
\indent We investigated in detail the presence of a perturber in Sect.~\ref{subsec:long_term} and no evidence is found. Assuming a negative periastron precession amplitude as observed of $\sim$-0.064$^\circ$d$^{-1}$ = -229.68$^{\prime\prime}$d$^{-1}$, through Eqs.~(12) and (16) in~\cite{Borkovits2011}, we can estimate the period of the body producing such a precession rate. We subtracted the general relativity contribution (calculated in Sect.~\ref{subsec:code} as 2.85$^{\prime\prime}$ d$^{-1}$) from these solutions and derived the orbital period of the perturber producing such an $\dot{\omega}$. For a circular orbit, a body with a mass in the range 0.1 M$_{\Earth}$ and 20 M$_{\Jupiter}$ would have an orbital period below 10 d, as shown in Fig.~\ref{fig:negative_wdot}.  Such a planet would be retrievable in the RV data, while no long-term (over a 10-day orbital period) companion in this mass range can produce a negative periastron precession. Therefore the solutions yielding negative $\dot{\omega}$ values can be discarded. \\
\noindent The results for the fitted parameters of the $\dot{\omega}$-positive solution (solution B) with lowest $\chi^2$ are shown in Table~\ref{tab:model_parameters}, along with their errorbars. With 99 RV points (outside the transit, 203 in total), 162 mid-transit and 11 mid-occultation points, and 19 free parameters, the reduced $\chi^2$ is 1.02. The observed value of the total contribution $\dot{\omega}$ from the DE-MCMC analysis of the data is (233$^{+25}_{-35}$)$^{\prime\prime}$d$^{-1}$ (see Table~\ref{tab:model_parameters}), corresponding to an apsidal motion period of $U$=$2 \pi / \dot{\omega}$ $\simeq$ 15 years. \\
Figure~\ref{fig:OC} shows the residuals of the transit and occultation mid-times with such a fit. Figure~\ref{fig:phase_folded_RV} shows the phase-folded RV curve and the residuals of our periastron precession fit.\\
\indent We subtract the General Relativity term and the tidal and rotational components relative to the star from the observed $\dot{\omega}$. What is left in Eq.~(\ref{eq:wdot_tot}) are the tidal and rotational components due to the planet. They are related to the interior of the planet and show clearly the dependence on Love number (see Eqs.~(\ref{eq:omega_dot_tidal}) and (\ref{eq:omega_dot_rot})). Then, $\Lovep$ can then be estimated by inverting these equations. \\
\indent The only unknown parameter is the rotational period $\Protp$ of the planet itself. Since there is no measurement of the rotational period of the planet, we can only assume a value for $\Protp$. The two extreme cases are presented when:
\begin{enumerate}
    \item[-] tidal forces lead to the disruption of the planet ($\Protp$=0.14 d), then we obtain 
    $\Lovep$= 0.07;\\ 
    \item[-] the planet is non-rotating ($\Porb/\Protp \rightarrow $ 0), then we obtain $\Lovep$=0.22. 
\end{enumerate}
The measured Love number of the planet must lie in between these two values. We can suppose that tidal forces synchronize the orbital and rotational periods quite fast. This means that $\Porb$ = $\Protp$. 
We can then derive $\Lovep$ from Eqs.~(\ref{eq:Rp}) and (\ref{eq:Sp}): $\Lovep$=0.20 $^{+0.02}_{-0.03}$, with a precision of 15$\%$.

\indent For comparison, the most up-to-date values of the tidal Love number for planets and satellites in the Solar System and exoplanets are listed in Table~\ref{tab:k2_SolarSystem_exop}. The value obtained for the fluid Love number of WASP-19Ab has the same order of magnitude as the other values derived for exoplanets. \\
\indent The lower the value, the more matter is condensed toward the center of the body. The  low Love number obtained, $\Lovep$=0.20, is an indication of a high-mass core. This has to be tested and verified with simulations of planetary interior structure through the equations of state, however, this is beyond the scope of this paper.\\
Moreover, if hydrostatic equilibrium is assumed (whereby the surface of the body can be described as an equipotential surface of the tidal and rotational potentials), the second-order fluid Love number $h_\mathrm{2p}$ can also be calculated. This assumption is close to reality if the body is a giant gaseous planet. In this case, $h_\mathrm{2,p}$=$\Lovep$+1 (\citealt{MunkMacDonald1960}), therefore $h_\mathrm{2,p}$=1.20 for synchronous rotation. This causes a small-shaped deformation that can be hard to observe photometrically \citep{Hellard2019}. \\

\noindent During the analysis, we noticed that the main contribution to the uncertainty of the fit is related to the transit fit. It is due to underestimated errorbars on the mid-transit times, which compromise the reliability of the results, increasing the $\chi^2$  of the fit. This was already suggested by~\cite{Mancini2013} regarding the 54 transits they analysed. Therefore, we increased the errorbars to 0.0005 d (approximately 43 s) if the reported uncertainties were lower than this threshold. \\
\indent We also underline the importance of acquiring new light curves. Precise transit and, mostly, occultation mid-times are needed to break the degeneracy between periastron precession and the change of orbital period due to for instance companions or orbital decay.\\


\subsection{Alternative explanations}
We investigated alternative scenarios to the presence of periastron precession. In particular, we searched for distant massive companions with the approach of~\cite{Knutson2014}. The long-term acceleration of the system was fitted by adding a $\gamma$ acceleration to the RV Eq.~(\ref{eq:rv_model}) as:
\begin{equation}
    V_\mathrm{acc} (t_\mathrm{j})=\dot{V_{\gamma}} \cdot (t_{\mathrm j}-t_0).
\end{equation}
We ran the DE-MCMC algorithm in the case of a circular and an eccentric orbit. 
The posterior distribution of $\dot{V_{\gamma}}$ has median of (7.37 $\pm$ 5.02)$\cdot 10^{-7}$m s$^{-2}$ for the circular case and (7.43 $\pm$ 4.89)$\cdot 10^{-7}$ m s$^{-2}$ for the eccentric one. Both results are compatible with zero and reproduce the results of~\cite{Knutson2014} of (7.52 $\pm$ 3.94) $\cdot 10^{-7}$ m s$^{-2}$. 
Our results increase the confidence level of a non-detection of long-term acceleration from~\cite{Knutson2014}'s 1.9$\sigma$ to 1.5$\sigma$.\\

\noindent We also investigated the tidal decay scenario, as it has been analysed and confirmed in the system WASP-12b (see respectively~\citealt{Patra2017, Yee2020, Turner2021, Wong2022}). 
The observed TTVs can be due to tidal decay or periastron precession. We fitted the RV semi-amplitude $K$ in Eq.~(\ref{eq:RVsemiamplitude}) as a time-varying parameter:
\begin{equation}
K(t_\mathrm{j})=K_\mathrm{0}+\dot{K}(t_\mathrm{j}-t_\mathrm{0}).
\end{equation}
Deriving Kepler's laws with respect to time, we can see that a change in the RV semi-amplitude results in an opposite change in the semi-major axis.\\
\indent For the circular orbit case, we obtained $\dot{K}$=(-2.77 $\pm$ 2.80)$\cdot$10$^{-8}$ m s$^{-2}$ while for the eccentric case (-2.60 $\pm$2.73) $\cdot$10$^{-8}$ m s$^{-2}$. As~\cite{Rosario2022}, we did not find significant evidence of orbital decay. Both in the case of a long-term velocity and of tidal decay, the results are compatible with zero and therefore considered as constant. \\

\begin{table*}[]
    \centering
    \resizebox{0.97\textwidth}{!}{
    \begin{tabular}{l c  l l}
        \hline \hline
        {\bf Body} & {\bf Fluid Love number} & {\bf Method} & {\bf Reference} \\
        &  {\bf $\Lovep$} & & \\
        \hline
         WASP-19Ab & 0.20 $^{+0.02}_{-0.03}$ & RVs and TTVs: periastron precession & This work \\
        WASP-18Ab &  0.62$^{+0.55}_{-0.19}$ & RVs and TTVs: periastron precession &~\cite{Csizmadia2019} \\
        WASP-103b$^\dagger$ & 0.59 $^{+0.23}_{-0.27}$ & LCs: shape deformation &~\cite{Barros2022} \\ 
        WASP-121b$^\dagger$ & 0.39 $_{-0.81}^{+0.71}$ & LCs: shape deformation &~\cite{Hellard2020}\\ 
        Jupiter & 0.565 $\pm$ 0.006 & Juno: Doppler shift measurements &  ~\cite{Durante2020} \\ 
        Saturn & 0.390 $\pm$ 0.024&  Cassini: gravitational perturbations on Enceladus, Tethys, Dione and Rhea & ~\cite{Lainey2017} \\ 
        Enceladus & 0.9896 $\pm 0.0103$ & Cassini: quadrupole gravity field and its hemispherical asymmetry &~\cite{Taubner2016}\\ 
        Neptune & 0.392 & Orbital perturbations on Triton and Nereid &~\cite{Jacobson2009} \\ 
        
        \hline
        {\bf Body} & {\bf Tidal Love number} & {\bf Method} & {\bf Reference} \\
        \hline
        HAT-P-13b & 0.31 $^{+0.08}_{-0.05}$ & RVs and TTVs: $\dot{e}$ and $\dot{\omega}$ &~\cite{Buhler2016} \\ 
        \hspace{0.5cm} " " & 0.81 $\pm$ 0.10 & RVs and TTVs: $\dot{e}$ and $\dot{\omega}$ &~\cite{Hardy2017}\\ 
        Mercury & 0.53 $\pm$ 0.03 & MESSENGER: Topography &~\cite{Konopliv2020} \\  
        Venus & 0.295 $\pm$ 0.066 & Magellan and Pioneer Venus Orbiter: Doppler tracking &~\cite{KonoplivYoder1996} \\ 
        Moon & 0.02416 $\pm$ 0.00022 & GRAIL and LLR: mapping of the gravitational field  &~\cite{Williams2014} \\ 
        Mars & 0.183 $\pm$ 0.009 & Mars Express: Doppler tracking and characterization of the orbits of Phobos and Deimos &~\cite{JacobsonLainey2014} \\ 
        Titan & 0.637 $\pm$ 0.112 & Cassini: response to Saturn's variable tidal field & ~\cite{Iess2012} \\ 
        
       
    \end{tabular}}
    \caption{Fluid and tidal Love numbers measured for Solar System planets and minor bodies and exoplanets. In the first half of the table, we show fluid Love numbers, which are comparable with our result.
    In the second half, we report tidal Love numbers. They refer to different quantities, so they are not comparable, except at longer periods when elastic stresses relax. 
    \newline $^\dagger$ Calculated from $h_2$ assuming hydrostatic equilibrium 
    %
    }
    \label{tab:k2_SolarSystem_exop}
\end{table*}

\section{Conclusions}
\label{sec:conclusions}

In this study, we review the theory on tidal interaction, as described by~\cite{Kopal1978} and reported with modern notation. 
Using assumptions on the Love number of the star, based on well-established theoretical models of its evolution and interior structure, the apsidal motion rate allows us to constrain the Love number $\Lovep$ of the planet.
We applied the results to a case study, namely, the system WASP-19Ab, by using archival mid-transit times, occultations, and radial velocity data, including previously unpublished RV observations made by HARPS. This dataset covers twelve years in total. First, we ruled out previous claims of additional planetary companions in the system as possible perturbers in the radial velocity curve that could hide the small contribution of the periastron precession motion. Moreover, we investigated the presence of a long-term acceleration in the RV dataset and of tidal decay. Both these scenarios have been rejected with respect to being compatible with a zero trend. Then, we used a model that accounts for an eccentric orbit and apsidal motion in order to fit the data. We detected the presence of apsidal motion in the planetary orbit and linked it to the Love number of the planet. By assuming synchronous rotation for the planet, we derived $\Lovep$=0.20$^{+0.02}_{-0.03}$. 
\indent The planetary fluid Love number is constrained in this system for the first time. There are few systems for which it has been measured. We take the opportunity to compare WASP-19Ab to them. Table~\ref{tab:k2_SolarSystem_exop} shows the fluid and tidal Love numbers of Solar system planets and bodies and exoplanets which are reported in the literature.  The derived value of the fluid Love number $\Lovep$ of WASP-19Ab is on the same order of magnitude as the one of other Jupiter-like planets (WASP-4b, WASP-18Ab, WASP-103b and WASP-121b). A low value suggests the presence of a dense core in the planet, similar to the case of Saturn, Neptune, and WASP-121b. Note that the Love numbers of WASP-12b and WASP-4b (derived by~\citealt{Campo2011} and~\citealt{Bouma2019}, respectively) are intentionally left out of the table. In the first case, the TTVs might be explained by tidal decay (\cite{Yee2020} and reference therein). In the second case, the presence of a second, long-period planet in the system (\citealt{Turner2022})  further complicates the TTV analysis. The results of~\cite{Bouma2020} were discussed further by~\cite{Harre2023} in the light of this second planet, making the determination of $\Lovep$ uncertain.\\
\indent No general conclusion on planetary physics can be drawn from this set of measured Love numbers because the sample is too small to be representative of the existing exoplanet population. We need observations of the Love number of more exoplanets.\\


\noindent We also characterize, for the first time, the stellar multiplicity of the system. 
We find a stellar companion with an apparent separation of 68$^{\prime\prime}$, corresponding to a physical distance of almost 20\,000 AU, from the system under analysis. However, its expected effect on the RV of WASP-19A is negligible. \\

\noindent {\it Acknowledgments} We acknowledge the support of DFG Research Unit 2440: "Matter Under Planetary Interior Conditions: High Pressure, Planetary, and Plasma Physics" and of DFG grants RA 714/14-1 within the DFG Schwerpunkt SPP 1992, Exploring the Diversity of Extrasolar Planets. L. M. B. would also like to thank Carl Knight for providing us with two transits of WASP-19Ab observed by the Ngileah Observatory, New Zealand, in May 2022. L. M. B. would also like to acknowledge H. Hu\ss mann and F. Sohl for the useful discussion on the different kinds of Love numbers.\\
\indent The work is based on observations collected at the European Southern Observatory under ESO programme 0104.C-0849.\\
This research has made use of the NASA Exoplanet Archive, which is operated by the California Institute of Technology, under contract with the National Aeronautics and Space Administration under the Exoplanet Exploration Program.\\
This work presents results from the European Space Agency (ESA) space mission $Gaia$ (\href{https://www.cosmos.esa.int/gaia}{https://www.cosmos.esa.int/gaia}). $Gaia$ data are being processed by the $Gaia$ Data Processing and Analysis Consortium (DPAC) (\href{DPAC, https://www.cosmos.esa.int/web/gaia/dpac/consortium}{DPAC, https://www.cosmos.esa.int/web/gaia/dpac/consortium}). Funding for the DPAC is provided by national institutions, in particular the institutions participating in the Gaia MultiLateral Agreement (MLA).\\

\nocite{*}
\bibliographystyle{aa} 
\bibliography{aa.bib} 

\newpage

\setcounter{figure}{0}     
  
\renewcommand\thefigure{A.\arabic{figure}}

\begin{appendix}
\section{Equations of apsidal motion}
\label{app:apsidal_motion}
\begin{figure}
    \centering
    \includegraphics[width=0.5\textwidth]{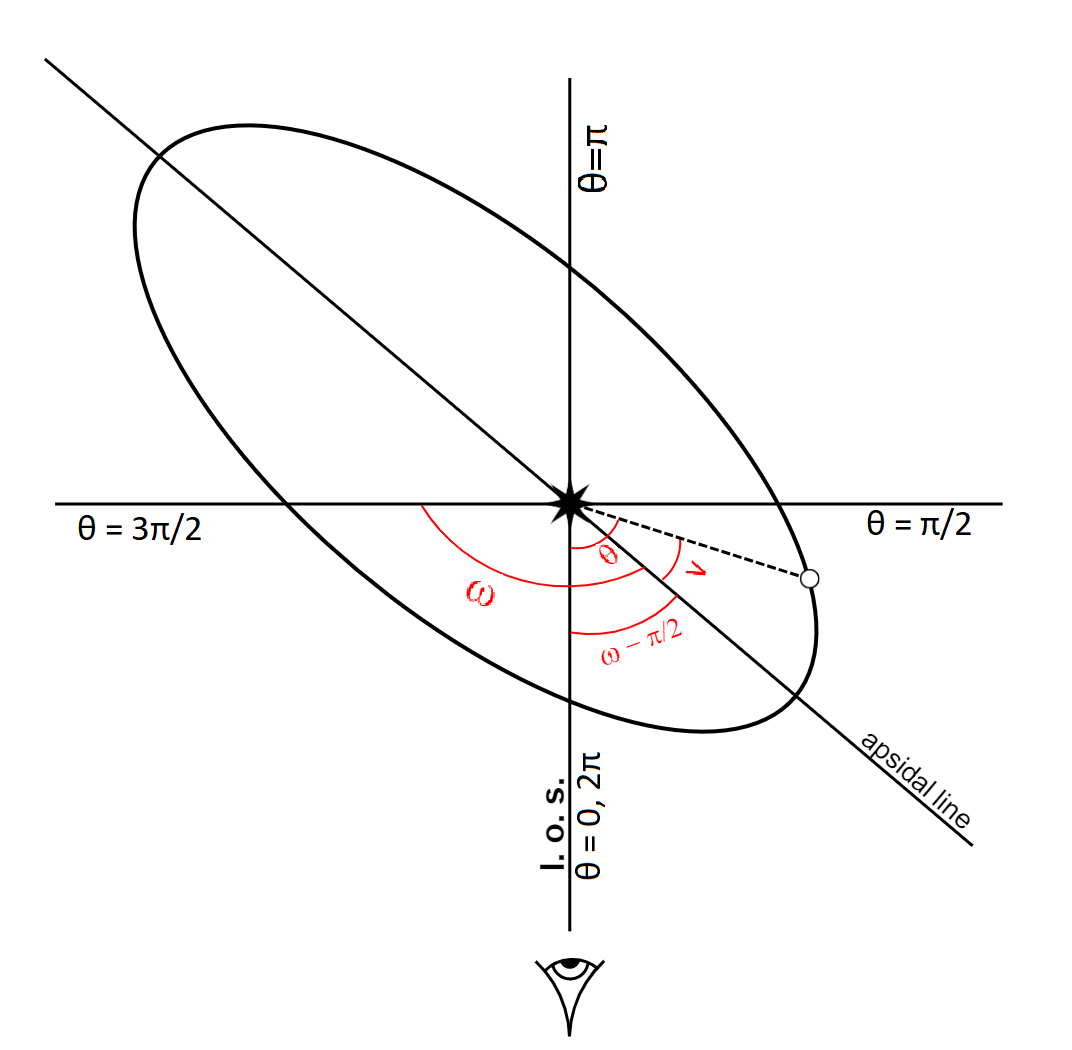}
    \caption{Geometry of the orbital system in the simple case of orbital inclination of $i$ =90$^\circ$. The star symbol and the open dot denote the positions of the star and the planet, respectively; $\omega$ is the angle of periastron, $\theta$ is the phase from superior conjunction, and $v$ is the true anomaly; $l.o.s.$ denotes the line of sight. In this configuration, $\theta$ equals 0 for transits and $\pi$ for occultations.}
    \label{Fig:system}
\end{figure}

In this Appendix, we rewrite the equations derived by \citet{Kopal1959,Kopal1978} with modern notation.\\
\indent First, we define the stellar and planetary disturbing accelerations $\mathbb{R}_\mathrm{i}$ and $\mathbb{S}_\mathrm{i}$, where the subscript $i$ represents either the star ($i = \star$) or the planet ($i=p$). $\mathbb{R}_\mathrm{i}$ acts in the direction of the radius vector and $\mathbb{S}_\mathrm{i}$ in the plane of the orbit in the direction perpendicular to the radius vector and positive in the direction of motion. \\
\indent The radial perturbation accelerations can be expressed as:
\begin{equation}
    \mathbb{R}_{\star} = - \frac{3}{4} G (\Ms + \mpl) \frac{\mpl}{\Ms} \Loves \frac{\Rs^5} {r^4 \hspace{0.1cm} r_{\mathrm{\epsilon_{\star}}}^3} (3 \cos^2\epsilon_\mathrm{\star}-1),
\end{equation}
and
\begin{equation}
    \mathbb{R}_\mathrm{p} = - \frac{3}{4} G (\Ms + \mpl)\frac{\Ms}{\mpl} \Lovep \frac{\rp^5} {r^4 \hspace{0.1cm} r_\mathrm{{\epsilon_p}}^3} (3 \cos^2\epsilon_\mathrm{p}-1), \label{eq:Rp} 
\end{equation}
where $G$ is the Gravitational constant, $\Ms$ and $\mpl$ are the masses of the star and the planet, respectively, $\Rs$ and $\rp$ their radii and $\Loves$ and $\Lovep$ their second-order fluid Love number. \\
\indent Here, $r$ is the radius vector:
\begin{equation}
    r = \frac{a (1-e^2)}{1+e \cdot \cos v},
    \label{eq:radius_vector}
\end{equation}
where $a$ is the semi-major axis of the orbit, $e$ is the orbital eccentricity and $v$ is the true anomaly of the planetary orbit, as shown in Fig.~\ref{Fig:system}. \\
\indent A tidal lag in latitude arises if the spin axis of the two bodies is not perpendicular to the orbital plane and in general is small or even zero. On the other hand, a lag in longitude arises from asynchronism between spin and orbital rotation. The angles $\epsilon_{\star}$ and $\epsilon_\mathrm{p}$ represent the tidal lag of the star or the planet in astrocentric longitude (\citealt{Efroimsky2013}) as:
\begin{equation}
    \epsilon_\mathrm{i} = \frac{1}{2} \arcsin \left( - \frac{1}{Q_i} \sign{[\Porb/P_\mathrm{rot,i} - 1 ] }\right),
\end{equation}
where $\Porb$ is the orbital period and $P_\mathrm{rot,i}$ is the rotational period of the star or the planet. The function $sign$ takes the value of +1, 0 or -1, depending on the sign of the expression in the square brackets. 
$Q_\mathrm{i}$ is the tidal quality factor. A high $Q$ factor corresponds to a weak dissipation and vice-versa (see~\cite{Kaula1964} and reference therein). \\
\indent The radius vector, $r_\mathrm{\epsilon}$, is defined as:
\begin{equation}
r_\mathrm{\epsilon,i} = \frac{a (1-e^2)}{1+e \cdot \cos(v-\epsilon_\mathrm{i})}.
\end{equation}
\indent The difference between $r$ and $r_\mathrm{\epsilon_i}$ is due to a finite speed of progression of the radial tidal waves in viscous matter. The amplitude of tidal distortion is not maximum when the bodies are closest to each other, but at a later time. We consider a fluid case, which corresponds to $\epsilon_\mathrm{i}$=0.\\

\noindent In addition, we define the tangential perturbation accelerations:
\begin{equation}
    \mathbb{S}_{\star} = \frac{3}{4} G (\Ms + \mpl) \frac{\mpl}{\Ms} \Loves \frac{{\Rs}^5}{r^4 r_{\epsilon_{\star}}^3} \sin(2\epsilon_{\star}),   
\end{equation}
for the star and
\begin{equation}
    \mathbb{S}_\mathrm{p} =\frac{3}{4} G (\Ms + \mpl)\frac{\Ms}{\mpl} \Lovep \frac{\rp^5}{r^4 r_\mathrm{{\epsilon_p}}^3} \sin(2\epsilon_\mathrm{p}), 
    \label{eq:Sp}  
\end{equation}
for the planet. Finally, $\mathbb{R}$ = $\mathbb{R}_\mathrm{\star}$ + $\mathbb{R}_\mathrm{p}$ and $\mathbb{S}$ = $\mathbb{S}_\mathrm{\star}$ + $\mathbb{S}_\mathrm{p}$. \\

\begin{figure}[!]
    \centering
    \includegraphics[width=0.49\textwidth]{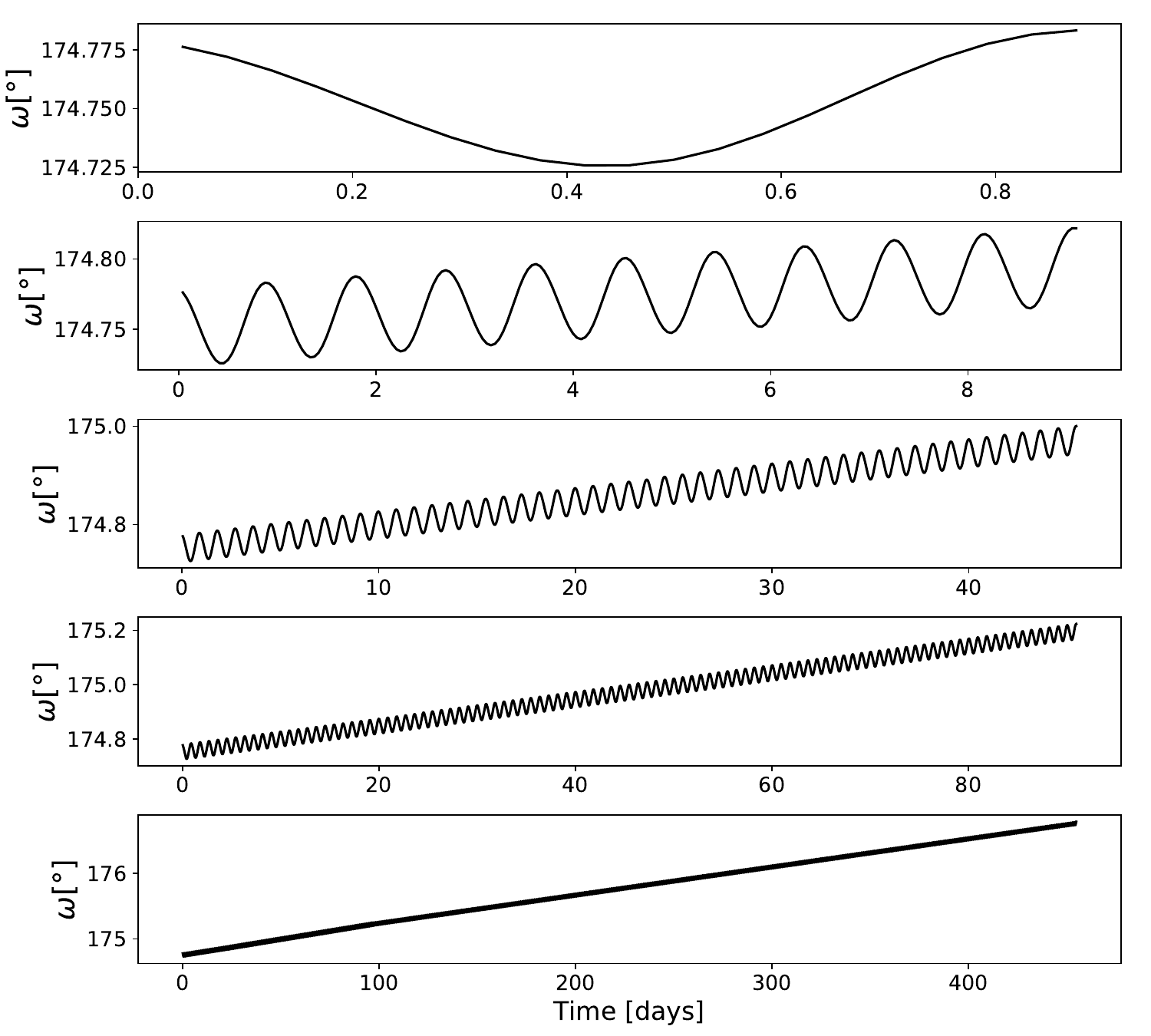}
    \caption{Temporal change of the osculating orbital element $\omega(t)$ 
    due to tidal and rotational potentials, as described by Eq.~(\ref{eq:domega/dt}), assuming the orbital properties of the system WASP-19 (see Sect.~\ref{subsec:systembackground}) and the Love number of WASP-18Ab \cite{Csizmadia2019}.}
    \label{fig:wdot}
\end{figure}

\noindent There is a third perturbing acceleration component beyond $\mathbb{S}$ and $\mathbb{R}$. It is normal to the plane of the orbit and positive toward the north pole. However, its contribution is zero as it is proportional to the latitude component of the angle of tidal lag. It can arise only if the orbital plane is inclined with respect to the equatorial plane of the body (star or planet). We neglect this term because the Rossiter-McLaughlin measurements of WASP-19A show an equatorial orbit (see, e.g.,~\citealt{Sedaghati2021}). \\
\indent With the introduction of the terms, $\mathbb{S}$ and $\mathbb{R}$, the perturbation equations can be easily written following \citet{Kopal1959,Kopal1978}.
The temporal change of the argument of periastron measured from the ascending node is:
\begin{equation}
\label{eq:domega/dt}
    \frac{d\omega}{dt} = \frac{\sqrt{1-e^2}}{n a e} \left[- \mathbb{R} \hspace{0.1cm} \cos v + \mathbb{S} \left(1+\frac{r}{a(1-e^2)} \right) \sin v \right],
\end{equation}
where $n$ is the mean motion that can be defined through Kepler's law or through the orbital period $\Porb$. \\
\indent Figure~\ref{fig:wdot} shows the effect of the periodical and secular perturbations on $\omega$, as in Eq.~(\ref{eq:domega/dt}). The plots were made setting the physical and orbital parameters to those of the exoplanetary system WASP-19Ab (see Sect.~\ref{sec:system}) with the Love number of WASP-18Ab \citep{Csizmadia2019} as a first guess. We present how it evolves over 1, 10, 50, 100, and 500 orbits. 

\section{Planetary eccentricity driven by WASP-19B}
\label{app:eccentricity_prove}
Here, we show that a companion star on a very eccentric orbit is able to cause an eccentric orbit of the planet in the WASP-19 system.

\cite{Borkovits2011} gave analytic expressions for the variations of the orbital elements up to the sixth order of the inner eccentricity for any arbitrary orientation of the orbits. However, we do not know the spatial orientation of the orbit of the companion star B nor the longitude of the node of the planet b. Therefore, we assume co-planar orbits for the planet and star B, and we also assume that their semi-major axes are aligned. In this specific case, the integration of Eq. (9) of~\cite{Borkovits2011} gives the simple expression for the eccentricity of the planet.
\begin{equation}
    e_b = \sqrt{\frac{e^{A_L/2 \cdot f(v_B)}}{1 + e^{A_L/2 \cdot f(v_B)}}}
    \label{eq:ecc_b}
,\end{equation}
where $f(v_B)$ is
\begin{equation}
    f(v_B) = \frac{1}{2} \cos 2v_B + \frac{4 e_B}{3} \cos^3 v_B, 
\end{equation}
and, $A_L,$
\begin{equation}
    A_L = \frac{15}{8} \frac{P_b}{P_B} \frac{M_B}{M_A + M_B + M_b} \frac{1}{(1-e_B)^{3/2}}.
\end{equation}
Here $v_B$ is the true anomaly of star B and it holds the time-dependence, $P$ and $M$ are the orbital periods around star A and the masses of the bodies and $e$ is the eccentricity. Indices $A$ and $B$ stand for stars A and B respectively, while index $b$ denotes the planet.

The lower period limit of star B around star A can be estimated assuming that $P_B$ is minimum if star B is in apoastron. In this case, the observed separation between stars A and B corresponds to a distance equal to $a_B (1+e_B)$. For different values of $e_B$, we can determine the semi-major axis that yields the orbital period, $P_B$, via Kepler's third law. In case of $e_B = 0.995$ $a_B \approx 10^4$ AU and $P_B \approx 1 $ Myrs (note: the periastron point is still located at 50 AU from star A).

Since we do not know the exact orbit of star B, we just estimate the planetary eccentricity. Taking the Taylor series of the exponential function in Eq.~(\ref{eq:ecc_b}) and neglecting the time dependence, we obtain
\begin{equation}
    e_b \approx \sqrt{A_L / 2 \times \left(\frac{1}{2} + \frac{4e_B}{3}\right)}.
\end{equation}
Using the aforementioned numerical values and assuming $M_B \sim 0.4 M_\odot$, we get $e_b \approx 0.0011$. 
This calculation leads to an order of magnitude estimation of whether any  (or a specific) combination of orbital elements can be responsible for the observed eccentricity. A deviation from this estimate can be due to the unknown spatial orientation of the different orbits.

\newpage
\setcounter{table}{0}
\renewcommand\thetable{C.\arabic{table}} 

\section{Tables of transits and occultations}
\label{app:TROCCdata}

\begin{table*}[]
\small
\resizebox{\textwidth}{!}{%
    \centering
    \begin{tabular}{l l l l l}
    \hline
    \hline
    {\bf {\large Cycle number}} &{\bf {\large T$_{mid}$ }}[days] & {\bf  {\large Uncertainty}} [days] & Observations & {\bf {\large Reference }}\\
N & $\BJDTDB$ - 2\,450\,000 & & & \\
\hline
0 & 4775.33720 & 0.00150 & WASP-South, 2.0m - Faulkes Telescope South (FTS) &~\cite{Hebb2010}\\ 
2 & 4776.91566 & 0.00019 & HAWK-I (VLT)  &~\cite{Anderson2010} \\ 
53 & 4817.14633 & 0.00021 & 2.0m - Faulkes
Telescope South (FTS) &~\cite{Lendl2013}, from~\cite{Hebb2010} \\ 
499 & 5168.96879 & 0.00009 & EFOSC on NTT, WASP-South and~\cite{Hebb2010}'s LCs &~\cite{Hellier2011} \\
517 & 5183.167890 & 0.000068 & IRAC on Spitzer &~\cite{Anderson2013} \\ 
538 & 5199.73343 & 0.000134 & 0.25m - Observatorio Astronómico Beta Orionis$^{\dagger 1}$ & F. Tifner (TRESCA)\\ 
604 & 5251.79657 & 0.00014 & EFOSC2 on NTT &~\cite{Tregloan-Reed2013} \\
605 & 5252.58544 & 0.00010 & EFOSC2 on NTT &~\cite{Tregloan-Reed2013} \\
609 & 5255.74077 & 0.00012 & EFOSC2 on NTT &~\cite{Tregloan-Reed2013} \\
609 & 5255.741050 & 0.000145 & EFOSC on NTT &~\cite{Dragomir2011}, from~\cite{Hellier2011} \\
614 & 5259.68459 & 0.000359 & 0.40m - Observatorio Cerro Armazones$^{\dagger 2}$  & Colque (TRESCA) re-analysed by~\cite{Petrucci2020}\\
632 & 5273.88282 & 0.00062 & 0.25m - Rarotonga Observatory$^{\dagger 3}$ & P. Evans (TRESCA) \\
664 & 5299.12768 & 0.00055 & 0.24m - YSVP Observatory$^{\dagger 4}$ & I. Curtis (TRESCA) \\
687 & 5317.27076 & 0.00006 & ASTEP 400 Telescope &~\cite{Abe2013} \\ 
709 & 5334.62540 & 0.00021 & DFOSC on 1.54m - Danish Telescope  &~\cite{Mancini2013}\\
714 & 5338.56929 & 0.00023 & TRAPPIST &~\cite{Lendl2013} \\ 
733 & 5353.55659 & 0.00024 & DFOSC on 1.54m Danish Telescope  &~\cite{Mancini2013}\\
746 & 5363.81131 & 0.00041 & 0.30m - MAGS Observatory$^{\dagger 5}$ & G. Milne (TRESCA) \\
752 & 5368.54285 & 0.00212 & DFOSC on 1.54m - Danish Telescope  &~\cite{Mancini2013} \\
969 & 5539.72329 & 0.00030 & EulerCam on Euler-Swiss  &~\cite{Lendl2013}  \\ 
1007 & 5569.69827 & 0.00036 & TRAPPIST &~\cite{Lendl2013}\\ 
1021 & 5580.74124 & 0.00058 & 1.0m - CTIO$^{\dagger 8}$ &~\cite{Dragomir2011} \\
1024 & 5583.10979 & 0.00089 & 0.24m - YSVP Observatory$^{\dagger 4}$ & I. Curtis (TRESCA) \\
1026 & 5584.68693 & 0.00024 & TRAPPIST &~\cite{Lendl2013}  \\ 
1026 & 5584.68685 & 0.00019 & EulerCam on Euler-Swiss &~\cite{Lendl2013}  \\ 
1038 & 5594.15188 & 0.00168 & 0.30m - PEST$^{\dagger 6}$ & TG Tan re-analysed by ~\cite{Mancini2013} \\
1047 & 5601.25164 & 0.00071 & D0.30m - PEST$^{\dagger 6}$ & TG Tan (TRESCA) re-analysed by~\cite{Mancini2013} \\
1049 & 5602.83138 & 0.00046 & TRAPPIST &~\cite{Lendl2013}  \\ 
1052 & 5605.19414 & 0.00180 & 0.30m - PEST$^{\dagger 6}$ & TG Tan (TRESCA) re-analysed by~\cite{Mancini2013}\\
1054 & 5606.77464 & 0.00022 & EulerCam on Euler-Swiss &~\cite{Lendl2013} \\ 
1055 & 5607.56241 & 0.00033 & TRAPPIST &~\cite{Lendl2013} \\ 
1074 & 5622.55057 & 0.00026 & TRAPPIST &~\cite{Lendl2013} \\ 
1076 & 5624.12787 & 0.00142 & DFOSC on 1.54m Danish Telescope  &~\cite{Mancini2013} \\
1087 & 5632.80612 & 0.00025 & EulerCam on Euler-Swiss &~\cite{Lendl2013}  \\ 
1116 & 5655.68222 & 0.00045 & TRAPPIST &~\cite{Lendl2013} \\ 
1135 & 5670.66976 & 0.00064 & TRAPPIST &~\cite{Lendl2013} \\ 
1144 & 5677.77038 & 0.00195 & 0.30m - PEST$^{\dagger 6}$ & TG Tan (TRESCA) re-analyzed by~\cite{Mancini2013}\\
1158 & 5688.81201 & 0.00333 & 0.30m - PEST$^{\dagger 6}$ & TG Tan (TRESCA) re-analyzed by ~\cite{Mancini2013} \\
1159 & 5689.60276 & 0.00030 & DFOSC on 1.54m Danish Telescope  &~\cite{Mancini2013} \\
1163 & 5692.75674 & 0.00255 & 0.30m - PEST$^{\dagger 6}$ & TG Tan (TRESCA) re-analyzed by~\cite{Mancini2013} \\
1164 & 5693.54639 & 0.00013 & DFOSC on 1.54m Danish Telescope  &~\cite{Mancini2013} \\
1177 & 5703.79933 & 0.00411 & 0.30m - PEST$^{\dagger 6}$ & TG Tan (TRESCA) re-analyzed by~\cite{Mancini2013} \\
1178 & 5704.59078 & 0.00034 & DFOSC on 1.54m Danish Telescope  &~\cite{Mancini2013} \\
1183 & 5708.53495 & 0.00015 & DFOSC on 1.54m Danish Telescope  &~\cite{Mancini2013} \\
1228 & 5744.032190 & 0.000040 & WFC3 on HST &~\cite{Mandell2013} \\
1254 & 5764.54014 & 0.00125 & CTIO 1m$^{\dagger 8}$  &~\cite{CortesZuleta2020}, analyzed by TLCM \\
1270 & 5777.16366 & 0.00022 & IRAC on Spitzer &~\cite{Wong2016} \\
1283 & 5787.41863 & 0.00023 & IRAC on Spitzer &~\cite{Wong2016} \\
1409 & 5886.81234 & 0.00208 & 0.30m - PEST$^{\dagger 6}$ & TG Tan (TRESCA) re-analyzed by  ~\cite{Mancini2013}\\
1421 & 5896.27611 & 0.00210 & 0.30m - PEST$^{\dagger 6}$ & TG Tan (TRESCA) re-analyzed by ~\cite{Mancini2013}\\
1445 & 5915.20980 & 0.00065 & 0.30m - PEST$^{\dagger 6}$ & TG Tan (TRESCA) re-analyzed by ~\cite{Mancini2013}\\
1450 & 5919.15485 & 0.00103 & 0.30m - PEST$^{\dagger 6}$ & TG Tan (TRESCA) re-analyzed by ~\cite{Mancini2013}\\
1454 & 5922.30966 & 0.00555 & 0.30m - PEST$^{\dagger 6}$ & TG Tan (TRESCA) re-analyzed by ~\cite{Mancini2013}\\
1552 & 5999.616301 & 0.00007 & MMIRS on the Magellan II Telescope &~\cite{Bean2013} \\
1580 & 6021.703740 & 0.000085 & MMIRS on the Magellan II Telescope  &~\cite{Bean2013} \\
1595 & 6033.53645 & 0.00014 & GROND$^{\dagger 7}$  &~\cite{Mancini2013}\\
1595 & 6033.53651 & 0.00007 & GROND$^{\dagger 7}$  &~\cite{Mancini2013}\\
1595 & 6033.53643 & 0.00007 & GROND$^{\dagger 7}$  &~\cite{Mancini2013}\\
1595 & 6033.53652 & 0.00009 & GROND$^{\dagger 7}$ &~\cite{Mancini2013}\\
1590 & 6029.59251 & 0.00035 & EulerCam on Euler-Swiss &~\cite{Lendl2013} \\ 
1633 & 6063.51175 & 0.00030 & TRAPPIST &~\cite{Lendl2013} \\ 
1977 & 6334.87208 & 0.00053 & 0.25m - Rarotonga Observatory$^{\dagger 3}$ & P. Evans (TRESCA)\\
1996 & 6349.86037 & 0.00081 & 0.25m - Rarotonga Observatory$^{\dagger 3}$ & P. Evans (TRESCA)\\
2016 & 6365.63794 & 0.00010 & DFOSC on 1.54m Danish Telescope  &~\cite{CortesZuleta2020}, analyzed by TLCM \\ 
2048 & 6390.87994 & 0.00043 & 0.25m - Rarotonga Observatory$^{\dagger 3}$ & P. Evans (TRESCA) \\
2064 & 6403.50164 & 0.00022 &  DFOSC on 1.54m Danish Telescope &~\cite{CortesZuleta2020}, analyzed by TLCM\\
2465 & 6719.82642 & 0.00050 & 0.25m - Rarotonga Observatory$^{\dagger 3}$ & P. Evans (TRESCA) \\
2490 & 6739.551364 & 0.000046 & IMACS on Magellan I Telescope &~\cite{Espinoza2019} \\
2490 & 6739.54756 & 0.00029 & 0.30m - FRAM$^{\dagger 9}$ & M. Mašek, K. Hoňková, J. Juryšek (TRESCA)\\ 
\end{tabular}
    }
    \caption{Transits found in the literature: cycle number, referred to in ~\cite{Hebb2010} - mid-transit point in $\mathrm{BJD}_\mathrm{TDB}$, its error, details on the observations, and reference. The time unit of each mid-transit time was checked and uniformized to $\mathrm{BJD}_\mathrm{TDB}$. Some digits are scientifically insignificant, but are left the same as in the reference papers. 
    \newline $^{\dagger 1}$ Beta Orionis Observatory, Argentina
    \newline $^{\dagger 2}$ Cerro Armazones Observatory, Argentina
    \newline $^{\dagger 3}$ Rarotonga Observatory, Cook Islands, New Zealand
    \newline $^{\dagger 4}$ YSVP Observatory, South Australia, Australia 
    \newline $^{\dagger 5}$ Mount Albert Grammar School Observatory, New Zealand
    \newline $^{\dagger 6}$ Perth Exoplanet Survey Telescope, West Australia, Australia
    \newline $^{\dagger 7}$ Gamma Ray Burst Optical and Near-Infrared Detector on MPG/ESO, La Silla, Chile
    \newline $^{\dagger 8}$ Cerro Tololo Inter-American Observatory, Chile
    \newline $^{\dagger 9}$ F/Photometric Robotic Atmospheric Monitor telescope, Pierre Auger Observatory, Argentina,~\cite{Ebr2014} 
    }
    \label{Tab:LiteratureTR_appendix}
\end{table*}

\begin{table*}
\resizebox{\textwidth}{!}{%
    \centering
\begin{tabular}{l l l l l}
    \hline
    \hline
       {\bf {\large Cycle number}} &{\bf {\large T$_{mid}$ }}[days] & {\bf  {\large Uncertainty }} [days] & Observations & {\bf {\large Reference }}\\
       N & $\mathrm{BJD}_\mathrm{TDB}$ - 2\,450\,000 & & & \\
        \hline
2532 & 6772.67789 & 0.00052 & 
TRAPPIST-South &~\cite{Patra2020} \\
2537 & 6776.625805 & 0.000066 & IMACS on Magellan I Telescope &~\cite{Espinoza2019} \\
2792 & 6977.77653 & 0.00019 & FORS2 on VLT &~\cite{Sedaghati2015}\\
2930 & 7086.63573 & 0.00020 & DFOSC on 1.54m Danish Telescope  &~\cite{CortesZuleta2020}, analyzed by TLCM\\
3044 & 7176.565524 & 0.000094 & IMACS on Magellan I Telescope &~\cite{Espinoza2019} \\
3351 & 7418.736819 & 0.000234 & FORS2 on VLT &~\cite{Sedaghati2017} re-analyzed by~\cite{Petrucci2020}\\ 
3389 & 7448.712917 & 0.0000766 & FORS2 on VLT &~\cite{Sedaghati2017} re-analyzed by~\cite{Petrucci2020}\\
3446 & 7493.67653 & 0.00024 & DFOSC on 1.54m - Danish Telescope  &~\cite{CortesZuleta2020}, analyzed by TLCM \\
3477 & 7518.131796 & 0.000380 & 0.30m - PEST$^{\dagger 6}$ &~\cite{IvshinaWinn2022} from~\cite{Petrucci2020}, TG Tan (TRESCA) \\%
3513 & 7546.53025 & 0.00038 & TRAPPIST-South &~\cite{Patra2020} \\
3513 & 7546.52975 & 0.0002759 & 1.54m - EABA$^{\dagger 10}$ &~\cite{IvshinaWinn2022} from~\cite{Petrucci2020} \\
3518 & 7550.472701 & 0.0004477 & 1.54m - EABA$^{\dagger 10}$ &~\cite{IvshinaWinn2022} from~\cite{Petrucci2020} \\
3551 & 7576.504698 & 0.0007447 & 0.30m - FRAM$^{\dagger 9}$ & M. Mašek, K. Hoňková, J. Juryšek (TRESCA), \\
& & & & re-analyzed by~\cite{IvshinaWinn2022} from~\cite{Petrucci2020} \\
3739 & 7724.807825 & 0.000828 & 2.15m - CASLEO$^{\dagger 11}$ & E. Fernández-Lajús, R. P. Di Sisto (TRESCA), re-analysed by~\cite{Petrucci2020} \\ 
3758 & 7739.794151 & 0.000537 & 1.54m - EABA$^{\dagger 10}$ &~\cite{Petrucci2020}\\ 
3806 & 7777.657426 & 0.001404 & 2.15m - CASLEO$^{\dagger 11}$ &~\cite{Petrucci2020}\\ 
3815 & 7784.758917 & 0.000176 & 1.54m - EABA$^{\dagger 10}$ &~\cite{Petrucci2020}\\ 
3825 & 7792.647526 & 0.000759 & 0.30m -FRAM$^{\dagger 9}$ & M. Mašek, K. Hoňková, J. Juryšek (TRESCA), re-analysed by~\cite{Petrucci2020}\\ 
3830 & 7796.591396 & 0.000316 & 0.30m - FRAM$^{\dagger 9}$ &  M. Mašek, K. Hoňková, J. Juryšek (TRESCA), re-analysed by~\cite{Petrucci2020} \\ 
3830 & 7796.592241 & 0.000061 & IMACS on Magellan I Telescope &~\cite{Espinoza2019} \\ 
3834 & 7799.74616 & 0.00013 & DFOSC on 1.54m - Danish Telescope  &~\cite{CortesZuleta2020}, analysed by TLCM \\
3896 & 7848.655592 & 0.000068 & IMACS on Magellan I Telescope &~\cite{Espinoza2019} \\ 
3901 & 7852.59809 & 0.00028 & DFOSC on 1.54m - Danish Telescope  &~\cite{CortesZuleta2020}, analysed by TLCM \\
3906 & 7856.543843 & 0.000111 & IMACS on Magellan I Telescope &~\cite{Espinoza2019} \\
3906 & 7856.542283 & 0.000524 & 0.30m - FRAM$^{\dagger 9}$ & M. Mašek, K. Hoňková, J. Juryšek (TRESCA), re-analysed by ~\cite{Petrucci2020} \\
4127 & 8030.87615 & 0.00213 & DFOSC on 1.54m - Danish Telescope  &~\cite{CortesZuleta2020}, analysed by TLCM \\
4189 & 8079.785673 & 0.0002142 & 1.54m - EABA$^{\dagger 10}$ &~\cite{Petrucci2020} \\
4227 & 8109.763837 & 0.001527 & 2.15m - CASLEO$^{\dagger 11}$ &~\cite{Petrucci2020} \\
4359 & 8213.887473 & 0.000243 & 0.30m - Ngileah Observatory$^{\dagger 12}$ &~\cite{Petrucci2020} from C. Knight (TRESCA) \\
4757 & 8527.845759 & 0.000453 & 2.15m - CASLEO$^{\dagger 11}$ &~\cite{Petrucci2020} \\
4778 & 8544.41145 & 0.0004479 & TESS &~\cite{IvshinaWinn2022}\\
4779 & 8545.19907 & 0.0003870 & TESS &~\cite{IvshinaWinn2022}\\
4780 & 8545.98902 & 0.0004050 & TESS &~\cite{IvshinaWinn2022}\\
4781 & 8546.77731 & 0.0004406 & TESS &~\cite{IvshinaWinn2022}\\
4782 & 8547.56698 & 0.0004338 & TESS &~\cite{IvshinaWinn2022}\\
4783 & 8548.35489 & 0.0004466 & TESS &~\cite{IvshinaWinn2022}\\
4784 & 8549.14427 & 0.0005019 & TESS &~\cite{IvshinaWinn2022}\\
4784 & 8549.144717 & 0.000534 &  0.32m - Bathurst Observatory$^{\dagger 13}$ &~\cite{Petrucci2020} from A. Wünsche (TRESCA) \\
4785 & 8549.93223 & 0.0004417 & TESS &~\cite{IvshinaWinn2022}\\
4786 & 8550.72120 & 0.0004841 & TESS &~\cite{IvshinaWinn2022}\\
4787 & 8551.50999 & 0.0004451 & TESS &~\cite{IvshinaWinn2022}\\
4788 & 8552.29907 & 0.0004410 & TESS &~\cite{IvshinaWinn2022}\\
4789 & 8553.08782 & 0.0005204 & TESS &~\cite{IvshinaWinn2022}\\
4790 & 8553.87694 & 0.0004185 & TESS &~\cite{IvshinaWinn2022}\\
4791 & 8554.66562 & 0.0004163 & TESS &~\cite{IvshinaWinn2022}\\
4792 & 8555.45470 & 0.00011 & TESS &~\cite{Wong2020} \\
4795 & 8557.82154 & 0.0005118 & TESS &~\cite{IvshinaWinn2022}\\
4796 & 8558.60951 & 0.0005083 & TESS &~\cite{IvshinaWinn2022}\\
4797 & 8559.39917 & 0.0004697 & TESS &~\cite{IvshinaWinn2022}\\
4798 & 8560.18675 & 0.0004762 & TESS &~\cite{IvshinaWinn2022}\\
4799 & 8560.97736 & 0.0004474 & TESS &~\cite{IvshinaWinn2022}\\
4800 & 8561.76555 & 0.0004190 & TESS &~\cite{IvshinaWinn2022}\\
4801 & 8562.55365 & 0.0004069 & TESS &~\cite{IvshinaWinn2022}\\
4802 & 8563.34279 & 0.0004283 & TESS &~\cite{IvshinaWinn2022}\\
4803 & 8564.13199 & 0.0004019 & TESS &~\cite{IvshinaWinn2022}\\
4804 & 8564.92114 & 0.0005022 & TESS &~\cite{IvshinaWinn2022}\\
4805 & 8565.70943 & 0.0004585 & TESS &~\cite{IvshinaWinn2022}\\
4806 & 8566.49870 & 0.0003981 & TESS &~\cite{IvshinaWinn2022}\\
4807 & 8567.28776 & 0.0004885 & TESS &~\cite{IvshinaWinn2022}\\
4808 & 8568.07637 & 0.0004006 & TESS &~\cite{IvshinaWinn2022}\\
5714 & 9282.76417 & 0.0003798 & TESS &~\cite{IvshinaWinn2022} \\
5715 & 9283.55339 & 0.0004562 & TESS &~\cite{IvshinaWinn2022} \\
5716 & 9284.34253 & 0.0005261 & TESS &~\cite{IvshinaWinn2022} \\
5717 & 9285.13026 & 0.0004526 & TESS &~\cite{IvshinaWinn2022} \\
5718 & 9285.91981 & 0.0004843 & TESS &~\cite{IvshinaWinn2022} \\
5719 & 9286.70858 & 0.0004234 & TESS &~\cite{IvshinaWinn2022} \\
5720 & 9287.49712 & 0.0004700 & TESS &~\cite{IvshinaWinn2022} \\
5721 & 9288.28588 & 0.0004688 & TESS &~\cite{IvshinaWinn2022} \\
5722 & 9289.07500 & 0.0004201 & TESS &~\cite{IvshinaWinn2022} \\
5723 & 9289.86356 & 0.0003895 & TESS &~\cite{IvshinaWinn2022} \\
5724 & 9290.65264 & 0.0004974 & TESS &~\cite{IvshinaWinn2022} \\
5725 & 9291.44097 & 0.0004855 & TESS &~\cite{IvshinaWinn2022} \\
5726 & 9292.23008 & 0.0004141 & TESS &~\cite{IvshinaWinn2022} \\

\end{tabular}
}

\caption{Continuation of Table~\ref{Tab:LiteratureTR_appendix}.
\newline $^{\dagger 9}$ F/Photometric Robotic Atmospheric Monitor telescope, Pierre Auger Observatory, Argentina
\newline $^{\dagger10}$ Estaci\'on Astrof\'isica de Bosque Alegre, Argentina
\newline $^{\dagger 11}$ Complejo Astron\'omico El Leoncito, Argentina
\newline $^{\dagger 12}$ Ngileah Observatory, New Zealand
\newline $^{\dagger 13}$ Bathurst Observatory, New South Wales, Australia
}
\end{table*}

\begin{table*}
    \resizebox{\textwidth}{!}{%
    \centering
\begin{tabular}{l l l l l}
    \hline
    \hline
       {\bf {\large Cycle number}} &{\bf {\large T$_{mid}$ }}[days] & {\bf  {\large Uncertainty }} [days] & Observations & {\bf {\large Reference }}\\
       N & $\mathrm{BJD}_\mathrm{TDB}$ - 2.450.000 & & & \\
        \hline
4824 & 8580.69724 & 0.00040 & TRAPPIST-South &~\cite{Patra2020} \\
4829 & 8584.641667 & 0.000268 & 1.54m - EABA$^{\dagger 9}$ &~\cite{Petrucci2020}\\
5730 & 9295.38506316 & 0.0005686 & TESS &~\cite{Rosario2022} \\
5731 & 9296.17539 & 0.0004310 & TESS &~\cite{IvshinaWinn2022} \\
5732 & 9296.96307 & 0.0004743 & TESS &~\cite{IvshinaWinn2022} \\
5733 & 9297.75141 & 0.0005115 & TESS &~\cite{IvshinaWinn2022} \\
5734 & 9298.54007 & 0.0004491 & TESS &~\cite{IvshinaWinn2022} \\
5735 & 9299.32928 & 0.0004109 & TESS &~\cite{IvshinaWinn2022} \\
5736 & 9300.11829 & 0.0004992 & TESS &~\cite{IvshinaWinn2022} \\
5737 & 9300.90761 & 0.0004934 & TESS &~\cite{IvshinaWinn2022} \\
5738 & 9301.69654 & 0.0004158 & TESS &~\cite{IvshinaWinn2022} \\
5739 & 9302.48453 & 0.0004775 & TESS &~\cite{IvshinaWinn2022} \\
5740 & 9303.27451 & 0.0004810 & TESS &~\cite{IvshinaWinn2022} \\
5741 & 9304.06354 & 0.0004313 & TESS &~\cite{IvshinaWinn2022} \\
5742 & 9304.85186 & 0.0004499 & TESS &~\cite{IvshinaWinn2022} \\
6258 & 9711.89359 & 0.00034 & 0.30m - Ngileah Observatory$^{\dagger 12}$ & C. Knight (TRESCA), analysed by TLCM \\
6272 & 9722.93627 & 0.00084 & 0.30m -  Ngileah Observatory$^{\dagger 12}$ & C. Knight (TRESCA), analysed by TLCM \\
6505 & 9906.734941 & 0.00058 & 0.43m - Deep Sky, Chile & Y. Jongen (TRESCA) \\
6141 & 9619.60147 & 0.00039 & 0.43m - Deep Sky, Chile & J.-P. Vignes (TRESCA) \\
6145 & 9622.756750 & 0.00042 & 0.43m - Deep Sky, Chile & J.-P. Vignes (TRESCA) \\
6557 & 9947.75209 & 0.00114 & 0.15m - Observatorio Vuelta por el Universo$^{\dagger 14}$ & M. Anzola (TRESCA) \\
    \end{tabular}
    }
    \caption{Continuation of Table~\ref{Tab:LiteratureTR_appendix}.
    \newline $^{\dagger 10}$ Estaci\'on Astrof\'isica de Bosque Alegre, Argentina
    \newline $^{\dagger 12}$ Ngileah Observatory, New Zealand
    \newline $^{\dagger 14}$ Observatorio Vuelta por el Universo, Argentina
    }
\end{table*}

\begin{table*}[]
\small
\resizebox{\textwidth}{!}{%
    \centering
    \begin{tabular}{l l l l l}
    \hline \hline
           {\bf {\large Cycle number}} &{\bf {\large T$_\mathrm{mid}$ }}[days] & {\bf  {\large Uncertainty }} [days] & Observations & {\bf {\large Reference }}\\
      N & $\BJDTDB$ - 2 450 000 & & \\
        \hline
        2 & 4777.313572 & $^{+0.00013}_{-0.00014}$ & HAWK-I (VLT) &~\cite{Anderson2010} \\ 
        185 & 4921.66849 & 0.00190 & HAWK-I (VLT)  &~\cite{Gibson2010} \\
        1018 & 5578.7684 & 0.0039 & ULTRACAM on NTT &~\cite{Burton2012} \\
        1199 & 5721.5508 & $^{+0.0028}_{-0.0031}$ & DFOSC on 1.54m - Danish Telescope &~\cite{Mancini2013}\\
        1269 & 5776.77020 & 0.00083 & IRAC on Spitzer &~\cite{Wong2016} \\
        1270 & 5777.55733 & 0.00092 & IRAC on Spitzer &~\cite{Wong2016} \\ 
        1282 & 5787.02383 & 0.00077 & IRAC on Spitzer &~\cite{Wong2016} \\
        1283 & 5787.81228 & 0.00076 & IRAC on Spitzer &~\cite{Wong2016} \\
        1549 & 5997.6440 & 0.001 & MMIRS on Magellan II telescope &~\cite{Bean2013} \\
        1582 & 6023.6763 & 0.001 & MMIRS on Magellan II telescope &~\cite{Bean2013} \\        
        \hline
\end{tabular}
    }
    \caption{Occultations found in the literature. The cycle number is referred to the previous transit (half phase before the occultation). Light travel time across the system (around 16.3 s) is subtracted from all mid-occultation times.
    }
    \label{Tab:LiteratureOCC}
\end{table*}

\newpage
\clearpage
\setcounter{table}{0}
\renewcommand\thetable{D.\arabic{table}} 
\section{Tables of RVs}
\label{app:RVdata}

\begin{table}[h]
    \centering
\resizebox{0.4\textwidth}{!}{%

\begin{tabular}{c c c c}
\hline
\hline
Instr. ID & $\mathrm{BJD}_\mathrm{TDB}$ & $RV_{obs}$ & Uncertainty  \\
& $-2~450~000.0$ & [km s$^{-1}$] & [km s$^{-1}$] \\
\hline
& & & \\
1 & 4616.46326 & 20.965 & 0.0218 \\
1 & 4623.46713 & 20.712 & 0.0289 \\
1 & 4624.46191 & 21.019 & 0.0220 \\
1 & 4652.46587 & 20.512 & 0.0210 \\
1 & 4653.46543 & 20.770 & 0.0231 \\
1 & 4654.46530 & 21.007 & 0.0219 \\
1 & 4656.47564 & 20.511 & 0.0193 \\
1 & 4657.46805 & 20.902 & 0.0386 \\
1 & 4658.46376 & 21.008 & 0.0302 \\
1 & 4660.46652 & 20.604 & 0.0400 \\
1 & 4661.46434 & 20.974 & 0.0217 \\
1 & 4662.46549 & 20.922 & 0.0191 \\
1 & 4663.46607 & 20.547 & 0.0188 \\
1 & 4664.46573 & 20.645 & 0.0362 \\
1 & 4665.46709 & 21.077 & 0.0305 \\
1 & 4827.74752 & 20.683 & 0.0188 \\
1 & 4832.74362 & 21.050 & 0.0242 \\
1 & 4833.66844 & 20.860 & 0.0212 \\
1 & 4834.67693 & 20.548 & 0.0191 \\
1 & 4837.67024 & 20.726 & 0.0227 \\
1 & 4838.68496 & 20.562 & 0.0244 \\
1 & 4839.70064 & 20.936 & 0.0185 \\
1 & 4890.61491 & 20.637 & 0.0187 \\
1 & 4894.68743 & 20.518 & 0.0290 \\
1 & 4895.69238 & 20.894 & 0.0179 \\
1 & 4896.66180 & 21.041 & 0.0145 \\
1 & 4897.65715 & 20.674 & 0.0175 \\
1 & 4898.66015 & 20.544 & 0.0197 \\
1 & 4939.53373 & 20.578 & 0.0164 \\
1 & 4940.52747 & 20.642 & 0.0154 \\
1 & 4941.53928 & 20.996 & 0.0156 \\
1 & 4942.52421 & 20.899 & 0.0169 \\
1 & 4943.53594 & 20.559 & 0.0191 \\
1 & 4944.52955 & 20.798 & 0.0155 \\
\hline
2 & 4972.4990 & 20.8874 & 0.0176 \\
2 & 4973.4656 & 20.6042 & 0.0182 \\
2 & 4999.4865 & 20.5571 & 0.0341 \\
\hline
3 & 5242.6892 & 20.7469 & 0.0036 \\
3 & 5242.8465 & 21.0203 & 0.0050 \\
3 & 5243.6598 & 21.0391 & 0.0042 \\
3 & 5243.8250 & 20.9547 & 0.0043 \\
3 & 5244.7229 & 20.7355 & 0.0031 \\
3 & 5247.7159 & 21.0254 & 0.0088 \\
3 & 5272.5453 & 20.5741 & 0.0044 \\
3 & 5272.7784 & 20.9778 & 0.0052 \\
3 & 5274.5336 & 21.0391 & 0.0073 \\
3 & 5274.6155 & 20.9318 & 0.0124 \\
3 & 5274.6232 & 20.9129 & 0.0112 \\
3 & 5274.6307 & 20.8819 & 0.0094 \\
3 & 5274.6383 & 20.8842 & 0.0088 \\
3 & 5274.6459 & 20.8711 & 0.0084 \\
3 & 5274.6533 & 20.8914 & 0.0086 \\
3 & 5274.6610 & 20.8473 & 0.0094 \\
3 & 5274.6687 & 20.8214 & 0.0100 \\
3 & 5274.6762 & 20.7819 & 0.0108 \\
3 & 5274.6838 & 20.7568 & 0.0118 \\
3 & 5274.6915 & 20.7315 & 0.0108 \\
3 & 5274.6991 & 20.7138 & 0.0103 \\
\hline
\end{tabular}
}
\caption{Radial velocity of WASP-19Ab system. The indices of the observations are the same as in Fig.~\ref{Fig:bjdtdb_RV} and  Sect.~\ref{sec:data}.}
\label{Tab:dataRV1}
\end{table}
\begin{table}[]
\centering
\resizebox{0.4\textwidth}{!}{%
\begin{tabular}{c c c c}
\hline
\hline
Instr. ID & $\mathrm{BJD}_\mathrm{TDB}$ & $RV_{obs}$ & Uncertainty \\
& $-2~450~000.0$ & [km s$^{-1}$] & [km s$^{-1}$] \\
\hline
& & & \\
3 & 5274.7066 & 20.7262 & 0.0106 \\
3 & 5274.7142 & 20.7093 & 0.0104 \\
3 & 5274.7218 & 20.7151 & 0.0118 \\
3 & 5274.7296 & 20.7056 & 0.0120 \\
3 & 5274.7371 & 20.6582 & 0.0116 \\
3 & 5274.7446 & 20.6687 & 0.0104 \\
3 & 5274.7522 & 20.6387 & 0.0119 \\
3 & 5274.7599 & 20.6452 & 0.0113 \\
3 & 5274.7675 & 20.6136 & 0.0105 \\
3 & 5274.7748 & 20.6079 & 0.0106 \\
3 & 5274.7826 & 20.5771 & 0.0123 \\
3 & 5274.8224 & 20.5707 & 0.0069 \\
3 & 5275.5229 & 20.6841 & 0.0052 \\
3 & 5275.7903 & 20.6841 & 0.0077 \\
3 & 5276.5154 & 20.5850 & 0.0062 \\
\hline
4 & 5338.47873 & 0.13439 & 0.01522 \\
4 & 5338.48498 & 0.11220 & 0.01214 \\
4 & 5338.49280 & 0.09687 & 0.01226 \\
4 & 5338.50068 & 0.06777 & 0.01130 \\
4 & 5338.50859 & 0.05913 & 0.01108 \\
4 & 5338.51623 & 0.04541 & 0.01068 \\
4 & 5338.52418 & 0.00975 & 0.01071 \\
4 & 5338.53215 & 0.00564 & 0.01140 \\
4 & 5338.53993 & 0.00808 & 0.01054 \\
4 & 5338.54787 & 0.06322 & 0.01216 \\
4 & 5338.55585 & -0.00877 & 0.01121 \\
4 & 5338.56351 & -0.01221 & 0.01281 \\
4 & 5338.57167 & -0.05589 & 0.01214 \\
\hline
5 & 6018.8117983 & -0.168663 & 0.003077 \\
5 & 6290.1405875 & -0.137606 & 0.003032 \\
5 & 6318.9215917 & 0.306274 & 0.002318 \\
\hline
6 & 8498.59068874 &     20.82646 &      0.00818 \\
6 & 8498.60180095 & 20.7993      &      0.00716 \\
6 & 8498.61279395 &     20.7785  &      0.00819 \\
6 & 8498.62459247 &     20.76231 &      0.00651 \\
6 & 8498.63454273 &     20.76321 &      0.0145 \\
6 & 8498.63972473 &     20.77791 &      0.01274 \\
6 & 8498.64464221 &     20.79349 &      0.01338 \\
6 & 8498.64956116 &     20.73967 &      0.01424 \\
6 & 8498.65436899 &     20.69906 &      0.01462  \\
6 & 8498.65942065 &     20.69273 &      0.01402 \\
6 & 8498.66430660 &     20.67341 &      0.01475 \\
6 & 8498.66926300 &     20.63671 &      0.01324 \\
6 & 8498.67428582 &     20.61285 &      0.01011 \\
6 & 8498.67907061 &     20.60608 &      0.00899  \\
6 & 8498.68407176 &     20.61570 &      0.00897  \\
6 & 8498.68899962 &     20.61782 &      0.00781  \\
6 & 8498.69387378 &     20.62464 &      0.00745  \\
6 & 8498.70087685 &     20.61316 &      0.00457  \\
6 & 8498.71044096 &     20.60040 &      0.00456  \\
6 & 8498.71956719 &     20.59894 &      0.00435  \\
6 & 8498.72920549 &     20.56458 &      0.00439  \\
6 & 8498.73862103 &     20.55429 &      0.00399  \\
6 & 8546.70109917 &     20.81404 &      0.00542  \\
6 & 8546.71810796 &     20.77644 &      0.00864  \\
6 & 8546.73102276 &     20.74880 &      0.00911  \\
6 & 8546.74715133 &     20.74357 &      0.00749  \\
6 & 8546.75677225 &     20.74175 &      0.01771 \\
6 & 8546.76244128 &     20.70465 &      0.00183  \\
6 & 8546.76773966 &     20.74182 &      0.02089 \\
6 & 8546.77334046 &     20.66661 &      0.01599 \\
6 & 8546.77859110 &     20.65585 &      0.01502 \\
\hline
\end{tabular}
}
\caption{Continuation of Table~\ref{Tab:dataRV1}.}
\label{Tab:dataRV2}
\end{table}

\begin{table}[]
\centering
\resizebox{0.4\textwidth}{!}{%
\begin{tabular}{c c c c}
\hline
\hline
Instr. ID & $\mathrm{BJD}_\mathrm{TDB}$ & $RV_{obs}$ & Uncertainty  \\
& $-2~450~000.0$ & [km s$^{-1}$] & [km s$^{-1}$] \\
\hline
& & & \\
6 & 8546.78395751 &     20.66390 &      0.01514 \\
6 & 8546.78909247 &     20.57827 &      0.02002 \\
6 & 8546.79475258 &     20.59739 &      0.01890 \\
6 & 8546.80036398 &     20.55623 &      0.02008 \\
6 & 8546.80553459 &     20.63033 &      0.01792 \\
6 & 8546.81107313 &     20.62332 &      0.01776 \\
6 & 8546.81638670 &     20.59751 &      0.01665 \\
6 & 8546.82327218 &     20.61843 &      0.01500 \\
6 & 8546.83186421 &     20.50721 &      0.01337 \\
6 & 8546.84018927 &     20.57332 &      0.01128 \\
6 & 8546.84855560 &     20.55209 &      0.01556 \\
6 & 8546.85730223 &     20.51568 &      0.01358 \\
6 & 8565.63006930 &     20.85050 &      0.00419 \\
6 & 8565.64050392 &     20.83263 &      0.00415  \\
6 & 8565.65068188 &     20.81295 &      0.00392  \\
6 & 8565.66089612 &     20.79194 &      0.00431  \\
6 & 8565.67121537 &     20.76472 &      0.00420  \\
6 & 8565.68117929 &     20.77662 &      0.00364  \\
6 & 8565.68892203 &     20.78292 &      0.00850  \\
6 & 8565.69398195 &     20.76429 &      0.00868  \\
6 & 8565.69899674 &     20.74261 &      0.00836  \\
6 & 8565.70399522 &     20.72989 &      0.00813  \\
6 & 8565.70895259 &     20.68067 &      0.00844  \\
6 & 8565.71404306 &     20.66203 &      0.00896  \\
6 & 8565.71903842 &     20.64335 &      0.00830  \\
6 & 8565.72397488 &     20.61127 &      0.00858  \\
6 & 8565.72902717 &     20.61737 &      0.00865  \\
6 & 8565.73403920 &     20.61288 &      0.00769  \\
6 & 8565.73901729 &     20.62037 &      0.00740  \\
6 & 8565.74406099 &     20.61360 &      0.00733  \\
6 & 8565.75100246 &     20.60935 &      0.00391  \\
6 & 8565.75992839 &     20.59612 &      0.00389  \\
6 & 8565.76886384 &     20.57339 &      0.00399  \\
6 & 8565.77776409 &     20.56269 &      0.00403  \\
6 & 8565.78359563 &     20.54291 &      0.01537 \\
\hline
7 & 8860.65868965 &     20.80633 &      0.00247  \\     
7 & 8860.66953784 &     20.78622 &      0.00219  \\
7 & 8860.68086427 &     20.77083 &      0.00224  \\
7 & 8860.69232959 &     20.75252 &      0.00205  \\
7 & 8860.70329540 &     20.73853 &      0.00201  \\
7 & 8860.71235224 &     20.76021 &      0.00306  \\
7 & 8860.71764990 &     20.75988 &      0.00329  \\
7 & 8860.72296232 &     20.74915 &      0.00298  \\
7 & 8860.72842374 &     20.72062 &      0.00296  \\
7 & 8860.73374936 &     20.68917 &      0.00305  \\
7 & 8860.73902683 &     20.65960 &      0.00289  \\
7 & 8860.74434671 &     20.62669 &      0.00289  \\
7 & 8860.74960873 &     20.60702 &      0.00300  \\
7 & 8860.75500031 &     20.59131 &      0.00293  \\
7 & 8860.76033814 &     20.59485 &      0.00285  \\
7 & 8860.76571960 &     20.60629 &      0.00280  \\
7 & 8860.77096557 &     20.60589 &      0.00288  \\
7 & 8860.77847463 &     20.59486 &      0.00195  \\
7 & 8860.78808536 &     20.57681 &      0.00200  \\
7 & 8860.79770303 &     20.56175 &      0.00208  \\
7 & 8860.80730381 &     20.54011 &      0.00213  \\
7 & 8860.81678339 &     20.52403 &      0.00211  \\
\hline
8 & 8885.81831 & 21.05010 & 0.00590\\
8 & 8886.63492 & 21.04724 & 0.00588 \\
8 & 8886.74637 & 20.91119 & 0.00495 \\
\hline
\end{tabular}
}
\caption{Continuation of Table~\ref{Tab:dataRV2}.}
\label{Tab:dataRV3}
\end{table}

\begin{table}[]
\centering
\resizebox{0.4\textwidth}{!}{%
\begin{tabular}{c c c c}
\hline
\hline
Instr. ID & $\mathrm{BJD}_\mathrm{TDB}$ & $RV_{obs}$ & Uncertainty  \\
& $-2~450~000.0$ & [km s$^{-1}$] & [km s$^{-1}$] \\
\hline
& & & \\
8 & 8887.70434 & 20.62644 & 0.01004\\
8 & 8887.78381 & 20.61007 & 0.00596\\
8 & 8888.61415 & 20.65507 & 0.00593 \\
8 & 8888.65498 & 20.69769 & 0.00576 \\
8 & 8889.67038 & 21.11750 & 0.00694 \\
8 & 8889.82412 & 21.04723 & 0.00663 \\
8 & 8893.66960 & 21.10599 & 0.01208 \\
8 & 8893.76426 & 21.04400 & 0.00549 \\
8 & 8893.83745 & 20.91112 & 0.00615 \\
8 & 8894.63550 & 20.90340 & 0.00821 \\
8 & 8894.69107 & 20.75424 & 0.00781 \\
8 & 8894.77058 & 20.62550 & 0.00732 \\
8 & 8897.68716 & 21.06761 & 0.00712 \\
8 & 8897.75751 & 20.97271 & 0.00668 \\
8 & 8899.85008 & 21.01457 & 0.00592 \\
\hline
\end{tabular}
}
\caption{Continuation of Table~\ref{Tab:dataRV3}.}
\label{Tab:dataRV4}
\end{table}

\end{appendix}

\end{document}